\documentclass[preprint2]{aastex}

\shortauthors{Vavrycuk}

\usepackage{graphicx}
\usepackage{amsmath}
\usepackage{esint}
\usepackage{tablefootnote}
\usepackage{threeparttable}

\usepackage{natbib} 

\begin{document} 

\title{Considering light-matter interactions in the Friedmann equations}

\author{V. Vavry\v cuk}

\affil{The Czech Academy of Sciences}
\affil{Bo\v cn\' i II 1401, 141 00 Praha 4}
\email{vv@ig.cas.cz}

\begin{abstract}
Recent observations indicate that the Universe is not transparent but partially opaque due to absorption of light by ambient cosmic dust. This implies that the Friedmann equations valid for the transparent universe must be modified for the opaque universe. The paper studies a  scenario when the opacity steeply rises with redshift. In this case, the light-matter interactions become important, because cosmic opacity produces radiation pressure that counterbalances gravitational forces. The presented theoretical model assumes the Universe expanding according to the standard FLRW metric but with the scale factor $a(t)$ depending on both types of forces: gravity as well as radiation pressure. The modified Friedmann equations predicts a cyclic expansion/contraction evolution of the Universe within a limited range of scale factors with no initial singularity. The model avoids dark energy and removes some other tensions of the standard cosmological model. The paper demonstrates that considering light-matter interactions in cosmic dynamics is crucial and can lead to new cosmological models essentially different from the standard $\Lambda$CDM model. This emphasizes necessity of new observations and studies of cosmic opacity and cosmic dust at high redshifts for more realistic modelling of the evolution of the Universe. 
\end{abstract}


\section{Introduction}

Dust is an important component of the interstellar medium (ISM) and intergalactic medium (IGM), which interacts with the stellar radiation. Dust grains absorb and scatter the starlight and reemit the absorbed energy at infrared, far-infrared and microwave wavelengths \citep{Mathis1990, Schlegel1998, Calzetti2000, Draine2003, Draine2011, Vavrycuk2018}. Since galaxies contain interstellar dust, they lose their transparency and become opaque. The most transparent galaxies are elliptical, while the spiral and irregular galaxies are more opaque, when more than 40\% of light of stars in galaxies is absorbed by the galactic dust \citep{Calzetti2001,  Holwerda2005b, Holwerda2007, Finkelman2008, Lisenfeld2008}. Similarly, the Universe is not transparent but partially opaque due to ambient cosmic dust. Absorption of light by intergalactic dust grains produces cosmic opacity, which is spatially dependent and varies with frequency and redshift \citep{Aguirre1999a, Aguirre2000, Corasaniti2006, Vavrycuk2018, Vavrycuk2019}. It can be measured by dust reddening being particularly appreciable at close distance from galaxies and in intracluster space \citep{Chelouche2007, Muller2008, Menard2010a}. \citet{Menard2010a} correlated the brightness of $\approx$85.000 quasars at $z > 1$ with the position of $24 \times 10^{6}$ galaxies at $z \approx 0.3$ derived from the Sloan Digital Sky Survey, and found an averaged intergalactic attenuation $A_V$ to about 0.03 mag. 

Alternatively, the cosmic opacity can be estimated from the hydrogen column densities of Lyman $\alpha$ (Ly$\alpha$) absorbers. Massive clouds with $N_\mathrm{HI} \approx 10^{21} \, \mathrm{cm}^{-2}$, called the damped Ly$\alpha$ absorbers (DLAs), are self-shielded and rich on cosmic dust. They are detected in galaxies as well as in the circumgalactic and intergalactic space \citep{Wolfe2005, Meiksin2009, Menard_Fukugita2012, Peek2015, Tumlinson2017}. Since a relation between the total hydrogen column density $N_\mathrm{H}$ and the color excess $E\left(B-V\right)$ is known: $N_\mathrm{H} /E\left(B-V\right) = 5.6-5.8 \times 10^{21} \, \mathrm{cm}^{-2} \, \mathrm{mag}^{-1} \,$ \citep{Bohlin1978,Rachford2002}, we get the ratio $N_\mathrm{H} / A_V \approx 1.87 \times 10^{21} \, \mathrm{cm}^{-2} \, \mathrm{mag}^{-1}$ for $R_V = 3.1$, which is a typical value for our Galaxy \citep{Cardelli1989, Mathis1990}. From observations of the mean cross-section density of DLAs, $\langle n \sigma \rangle =  \left(1.13 \pm 0.15 \right) \times 10^{-5} \, h \, \mathrm{Mpc}^{-1}$ \citep{Zwaan2005}, the characteristic column density of DLAs, $N_\mathrm{HI} \approx 10^{21} \, \mathrm{cm}^{-2}$, and the mean molecular hydrogen fraction in DLAs of about $0.4 - 0.6$  \citep[their table 8]{Rachford2002}, we obtain the cosmic opacity $\lambda_V \approx 1-2 \times 10^{-5} \, h \, \mathrm{Mpc}^{-1}$ at $z = 0$.

\begin{figure*}
\includegraphics[angle=0, width=16cm, trim=50 180 70 120, clip = true]{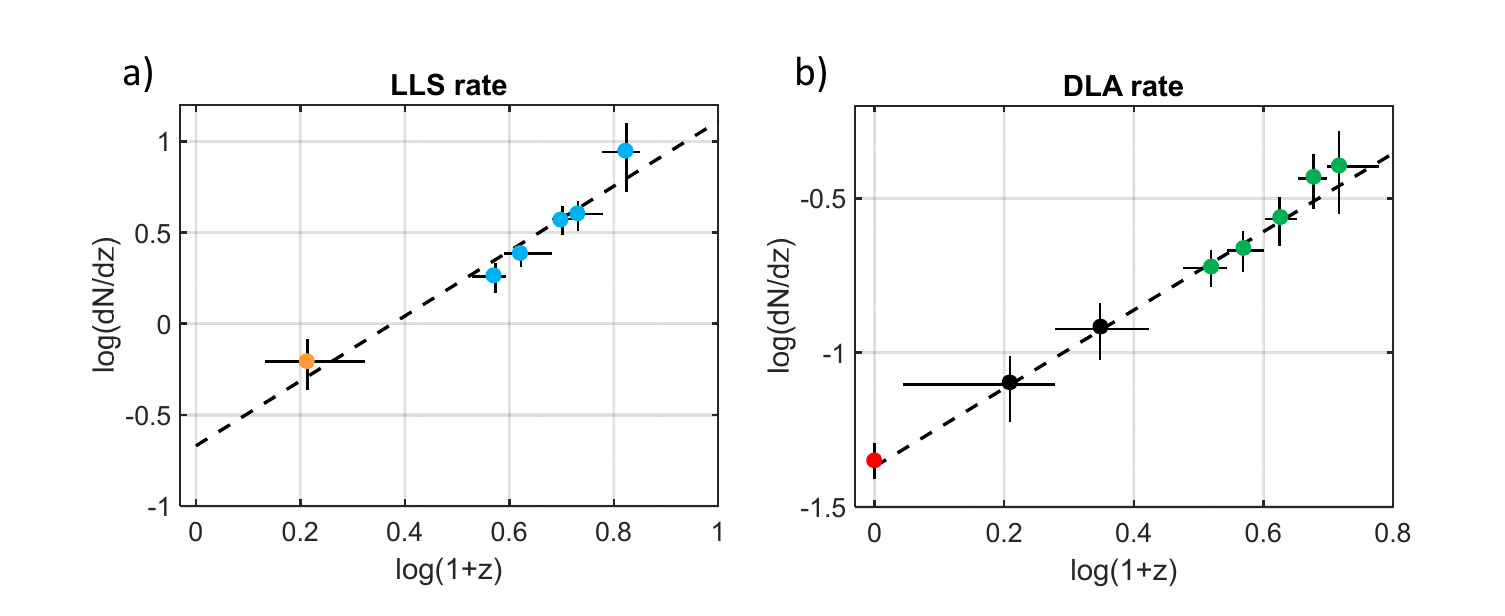}
\caption{
The incidence rate of the LLS (a) and DLA systems (b) as a function of redshift. The black dashed line - interpolation of observations. The observations are taken from \citet{Peroux2003} - orange dot, \citet{Songaila_Cowie2010} - cyan dots, \citet{Zwaan2005} - red dot, \citet{Rao2006} - black dots, and \citet{Prochaska_Herbert-Fort2004} - green dots. 
}
\label{fig:1}
\end{figure*}

The cosmic opacity is very low in the local Universe  \citep{Chelouche2007, Muller2008}, but it might steeply increase with redshift \citep{Menard2010a, Xie2015, Vavrycuk2017a}. Appreciable cosmic opacity at high redshift is documented by observations of (1) the evolution of the Ly$\alpha$ forest of absorption lines in quasar optical spectra, (2) the metallicity detected in the Ly$\alpha$ forest, and (3) emission spectra of high-redshift galaxies. In the Ly$\alpha$ forest studies, the evolution of massive Lyman-limit (LLS) and damped Lyman absorption (DLA) systems are, in particular, important, because they serve as reservoirs of dust \citep{Wolfe2005, Meiksin2009}. It has been shown that the incidence rate and the Gunn-Peterson optical depth of the LLS and DLA systems increase with redshift as $(1+z)^4$ or more for $z < 7$ \citep{Prochaska_Herbert-Fort2004, Fan2006b, Rao2006, Songaila_Cowie2010}, see Fig.~\ref{fig:1}. For higher $z$, the increase of the optical depth is even stronger. 

Another independent indication of dust at high redshifts is a weak or no evolution of metallicity with redshift. For example, observations of the C$_{\mathrm{IV}}$ absorbers do not show any visible redshift evolution over cosmic time suggesting that a large fraction of intergalactic metals may already have been in place at $z > 6$ \citep{Songaila2001, Pettini2003, Ryan-Weber2006}. In addition, the presence of dust in the high-redshift universe is documented  by observations of dusty galaxies even at $z > 7$ \citep{Watson2015, Laporte2017} and dusty halos around star-forming galaxies at $z = 5-7$ \citep{Fujimoto2019}. \citet{Zavala2015} measured a dust mass of $ \approx 10^7 M_\odot$ for a galaxy at $z \approx 9$. Since dust in high-redshift galaxies can efficiently be transported to halos due to galactic wind \citep{Aguirre1999a, Aguirre1999b} and radiation pressure \citep{Hirashita_Inoue2019}, the cosmic dust must be present even at redshifts $z > 7-9$. 

%

\begin{figure*}
\includegraphics[angle=0, width=17cm, trim=140 220 100 140, clip = true]{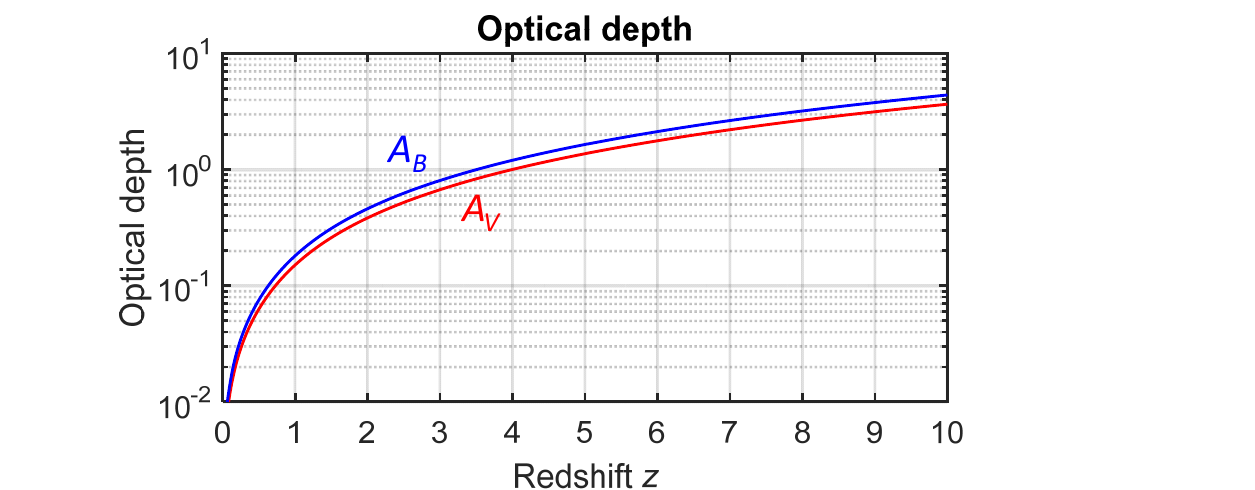}
\caption{
Optical depth of intergalactic space as a function of redshift. The extinction coefficient $R_V = A_V/(E(B-E)$ is assumed to be 5. $A_V$ - extinction at the visual band, $A_B$ - extinction at the B band. For details, see \citet{Vavrycuk2017a,Vavrycuk2018}.
}
\label{fig:2}
\end{figure*}

Since dust is traced mostly by reddening of galaxies and quasars at high redshifts, it is difficult to distinguish which portion of reddening is caused by dust present in a galaxy and by cosmic dust along the line of sight. \citet{Xie2015,Xie2016} studied dust extinction using spectra of $\approx$90.000 quasars from the SDSS DR7 quasar catalogue and tried to separate both the effects. They revealed that quasars have systematically redder UV continuum slopes at higher redshifts and estimated the extinction $A_V$ by cosmic dust of about $\approx 0.02 \, \mathrm{Gpc}^{-1}$. This value, however, strongly increases with redshift, because of increase of dust density due to the smaller volume of the Universe in the past \citep{Vavrycuk2017a,Vavrycuk2018}, see Fig.~\ref{fig:2}.

%
\begin{figure*}
\includegraphics[angle=0, width=16cm, trim=60 160 50 140, clip = true]{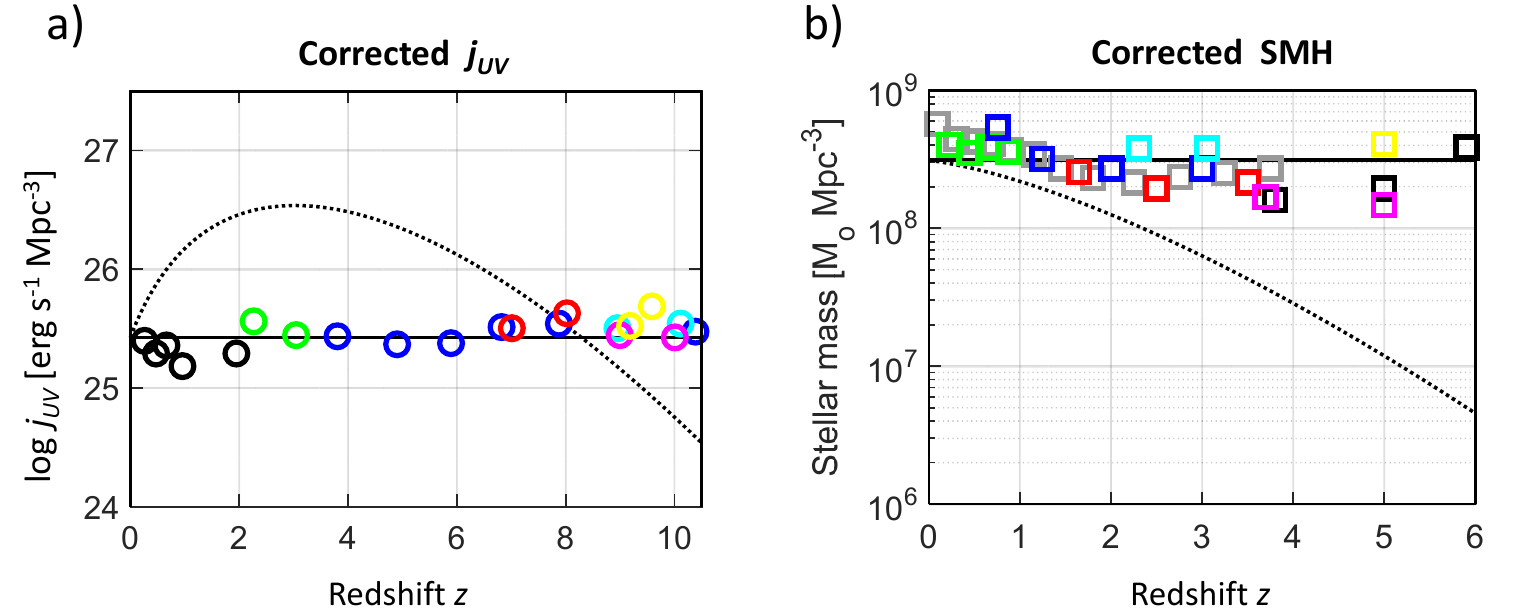}
\caption{
(a) The corrected comoving UV luminosity density $j_{UV}$ as a function of redshift after eliminating the effect of the cosmic opacity defined by $A_{UV}$ of $0.08 \,\mathrm{mag}\, h\, \mathrm{Gpc}^{-1}$. Observations are taken from \citet[black circles]{Schiminovich2005}, \citet[green circles]{Reddy2009}, \citet[blue circles]{Bouwens2014a}, \citet[red circles]{McLure2013}, \citet[magenta circles]{Ellis2013}, \citet[cyan circles]{Oesch2014}, and \citet[yellow circles]{Bouwens2014b}. The dotted line shows the apparent comoving luminosity density, when the bias produced by the cosmic opacity is not eliminated. (b) The comoving global stellar mass history (SMH) after eliminating the effect of the cosmic opacity defined by $A_{UV}$ of $0.08 \,\mathrm{mag}\, h\, \mathrm{Gpc}^{-1}$. 
The colour squares show observations reported by \citet[grey]{Perez_Gonzalez2008}, \citet[green]{Pozzetti2010}, \citet[blue]{Kajisawa2009}, \citet[red]{Marchesini2009}, \citet[cyan]{Reddy2012}, \citet[black]{Gonzalez2011}, \citet[magenta]{Lee2012}, and \citet[yellow]{Yabe2009}. The values are summarized in Table 2 of \citet{Madau_Dickinson2014}. The dotted line shows the apparent comoving SMH, when the bias produced by the cosmic opacity is not eliminated. For details, see \citet{Vavrycuk2018}.
}
\label{fig:3}
\end{figure*}
%

The fact that the Universe is not transparent but partially opaque might have fundamental cosmological consequences, because the commonly accepted cosmological model was developed for the transparent universe. Neglecting cosmic opacity produced by intergalactic dust may lead to distorting the observed evolution of the luminosity density and the global stellar mass density with redshift \citep{Vavrycuk2017a}. For example, a decrease of the luminosity density with redshift observed for $z > 2-3$ is commonly explained by darkness of the early Universe. However, this effect can just be an artefact of non-negligible opacity of IGM in the early Universe, when the light coming from high redshifts is attenuated \citep{Vavrycuk2017a}. Fig.~\ref{fig:3} shows that after eliminating the effect of the opacity from observations, the comoving luminosity density and global stellar mass is redshift independent. Note that physical origin of darkness of the early Universe discussed here is quite different from that of 'dark ages' in the Big Bang theory. While we study the cosmic opacity due to the presence of dust at redshifts $z < 25$ (dust temperature being less than 80 K), the dark ages epoch is produced by opaque plasma at redshifts $z > 1100$ (plasma temperature being $\sim 10^{9}$ K).

Non-zero cosmic opacity may invalidate the interpretation of the Type Ia supernova (SNe Ia) dimming as a result of dark energy and the accelerating expansion of the Universe \citep{Aguirre1999a, Aguirre1999b, Aguirre_Haiman2000, Menard2010b}. According to \citet{Vavrycuk2019} and \citet{Vavrycuk_Kroupa2020}, cosmic opacity 
$\lambda_B \approx 0.08-0.10 \, \mathrm{Gpc}^{-1}$ fits the Type Ia supernova observations with no need to introduce the accelerated expansion. In addition, cosmic dust can partly or fully produce the cosmic microwave background (CMB) \citep{Wright1982, Bond1991, Narlikar2003}. For example, \citet{Vavrycuk2018} showed that thermal radiation of dust is capable to explain the spectrum, intensity and temperature of the CMB including the CMB temperature/polarization anisotropies. In this theory, the CMB temperature fluctuations are caused by fluctuations of the extragalactic background light (EBL) produced by galaxy clusters and voids in the Universe, and the CMB polarization anomalies originate in the polarized thermal emission of needle-shaped conducting dust grains, which are aligned by magnetic fields around large-scale structures such as clusters and voids.

If cosmic opacity and light-matter interactions are considered, the Friedmann equations in the current form are inadequate and must be modified. The radiation pressure, which is caused by absorption of photons by dust grains and acts against gravitational forces, must be incorporated. In this paper, I demonstrate that the radiation pressure due to light absorption is negligible at the present epoch, but it could be significantly stronger in the past epochs. Surprisingly, its rise with redshift could be so steep that it could even balance the gravitational forces at high redshifts and cause the expansion of the Universe. Based on numerical modelling and observations of basic cosmological parameters, I show that the modified Friedmann equations avoid the initial singularity and lead to a cyclic model of the Universe with expansion/contraction epochs within a limited range of scale factors. I estimate the maximum redshift of the Universe achieved in the past and the maximum scale factor of the Universe in future.

\section{Theory}

\subsection{Friedmann equations for the transparent universe}

The standard Friedmann equations for the pressureless fluid read \citep{Peacock1999,Ryden2016}
\begin{equation}\label{eq1}
{\left({\frac{\dot a}{a}}\right)}^2 = \frac{8\pi G}{3} \rho - \frac{k c^2}{a^2} + \frac{1}{3} \Lambda c^2 \,, 
\end{equation}
\begin{equation}\label{eq2}
\frac{\ddot{a}}{a} = -\frac{4\pi G}{3} \rho + \frac{1}{3} \Lambda c^2 \,, 
\end{equation}
where $a = R/R_0 = \left(1+z\right)^{-1}$ is the relative scale factor, $G$ is the gravitational constant, $\rho$ is the mass density, $k/a^2$ is the spatial curvature of the universe, $\Lambda$ is the cosmological constant, and $c$ is the speed of light. Considering mass density $\rho$ as a sum of matter and radiation contributions, we get
\begin{equation}\label{eq3}
\frac{8\pi G}{3} \rho =  H^2_0 \left[{\Omega_m a^{-3} + \Omega_r a^{-4}}\right] \,.
\end{equation}
Eq. (1) is then rewritten as
\begin{equation}\label{eq4}
H^2\left(a\right) = H^2_0 \left[{\Omega_m a^{-3} + \Omega_r a^{-4} + \Omega_\Lambda + \Omega_k a^{-2}}\right] \,,
\end{equation}
with the condition
\begin{equation}\label{eq5}
\Omega_m + \Omega_r + \Omega_\Lambda + \Omega_k = 1 \,, 
\end{equation}
where $H(a) = \dot{a}/a$ is the Hubble parameter, $H_0$ is the Hubble constant, and $\Omega_m$, $\Omega_r$ , $\Omega_\Lambda$ and $\Omega_k$ are the normalized matter, radiation, vacuum and curvature terms. Assuming  $\Omega_r = 0$ and $\Omega_k = 0$ in Eq. (4), we get the $\Lambda$CDM model 
\begin{equation}\label{eq6}
H^2\left(a\right) = H^2_0 \left[{\Omega_m a^{-3} +  \Omega_\Lambda }\right] \,,
\end{equation}
which describes a flat, matter-dominated universe. The universe is transparent, because any interaction of radiation with matter is neglected. The vacuum term $\Omega_\Lambda$ is called dark energy and it is responsible for the accelerating expansion of the Universe. The dark energy is introduced into Eqs (4-6) to fit the $\Lambda$CDM model with observations of the Type Ia supernova dimming.

\subsection{Light-matter interaction}

\begin{figure*}
\includegraphics[angle=0, width=16cm, trim=50 130 70 120, clip = true]{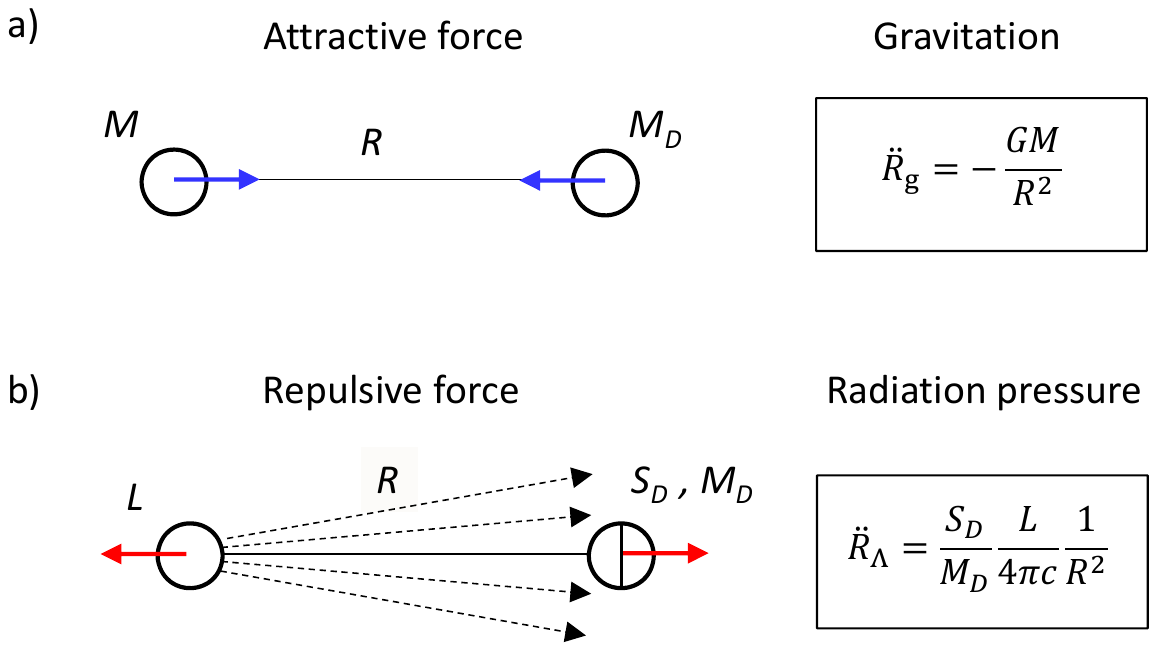}
\caption{
The scheme of gravitational forces (a) and radiation pressure (b) acting on dust grains. The blue and red arrows indicate a direction of the acting attractive and repulsive forces, respectively. The point source is characterized by mass $M$ and luminosity $L$. The dust grains have mass $M_D$ and the cross-section $S_D$. The radiation pressure caused by absorption of energy flux $I$ emitted by the light source with luminosity $L$ decreases with distance as $1/R^2$ similarly as the gravitational force.
}
\label{fig:4}
\end{figure*}
%

The basic drawback of the $\Lambda$CDM model is its assumption of transparency of the Universe and the neglect of the universe opacity caused by interaction of light with intergalactic dust. Absorption of light by cosmic dust produces radiation pressure acting against the gravity, but this pressure is ignored in the $\Lambda$CDM model.

Let us consider light emitted by a point source with mass $M$ and luminosity $L$ (in W) and absorbed by a dust grain with mass $M_D$, see Fig.~\ref{fig:4}. The light source produces the energy flux $I$ (in $\mathrm{W m}^{-2}$) and the radiation pressure $p_D$, which acts on the dust grain. The acceleration of the dust grain produced by the light source reads
\begin{equation}\label{eq7}
\ddot R_\Lambda = \frac{S_D}{M_D} p_D \,,
\end{equation}
where $S_D$ is the absorption cross-section of the grain. Since the radiation pressure $p_D$ is related to the energy flux $I$ and to the luminosity $L$ as
\begin{equation}\label{eq8}
p_D = \frac{I}{c} = \frac{L}{4\pi R^2 c} \,,
\end{equation}
we get
\begin{equation}\label{eq9}
\ddot R_\Lambda = \frac{S_D}{M_D} \frac{L}{4\pi c} \frac{1}{R^2} \,,
\end{equation}
where $R$ is the distance of the dust grain from the light source, and $c$ is the speed of light. The ratio $S_D/M_D$ in Eq. (9) can be expressed as
\begin{equation}\label{eq10}
\frac{S_D}{M_D} = \frac{3}{4} \frac{Q_\mathrm{abs}}{R_D \rho_D} = \kappa \,,
\end{equation}
where $S_D = Q_\mathrm{abs} \pi R^2_D$ is the absorption cross-section of the dust grain, $M_D = \frac{4}{3} \pi R^3_D \rho_D$ is the mass of the grain, $R_D$ is the grain radius, $Q_\mathrm{abs}$ is the grain absorption efficiency, $\rho_D$ is the specific mass density of grains, and $\kappa$ is the mass opacity (in $\mathrm{m}^2 \mathrm{kg}^{-1}$). Inserting Eq. (10) into Eq. (9), we write
\begin{equation}\label{eq11}
\ddot R_\Lambda = \frac{\kappa L}{4 \pi c} \frac{1}{R^2} \,.
\end{equation}

Comparing the radiation-absorption acceleration $\ddot R_\Lambda$ with the gravitational acceleration $\ddot R_g$  
\begin{equation}\label{eq12}
\ddot R_g = -\frac{G M}{R^2} \,,
\end{equation}
we see that both accelerations depend on distance from a source in the same way (as $1/R^2$). Consequently, the total acceleration of a dust grain is
\begin{equation}\label{eq13}
\ddot R = \ddot R_g + \ddot R_\Lambda = \frac{1}{R^2} \left(-G M + \frac{\kappa L}{4 \pi c} \right) \,.
\end{equation}

Dividing Eq. (13) by distance $R$ and substituting mass $M$ (in kg) and luminosity $L$ (in W) by mass density $\rho$ (in $\mathrm{kg m}^{-3}$) and luminosity density $j$ (in $\mathrm{W m}^{-3}$), we get
\begin{equation}\label{eq14}
\frac{\ddot R}{R} =  -\frac{4 \pi G}{3} \rho + \frac{\kappa j}{3 c} \,,
\end{equation}
and consequently, we obtain a generalized Poisson equation for the scalar potential $\Phi$, which involves potentials for both gravitational and radiation-absorption fields
\begin{equation}\label{eq15}
\Delta \Phi =  4 \pi G \rho - \frac{\kappa j}{c} \,.
\end{equation}
Equivalently
\begin{equation}\label{eq16}
\Delta \Phi =  4 \pi G \rho - \rho_\Lambda \,.
\end{equation}
where $\rho_\Lambda = \kappa j/c$ will be called the density of the radiation-absorption field.

\subsection{Friedmann equations for the opaque universe}

The generalized Poisson equation (16) implies that the radiation-absorption term is in many aspects similar to gravity; its effect is, however, opposite. Therefore, deriving the Friedmann equations for the opaque universe using general relativity will be analogous to that for the transparent universe. The only difference is that we have to introduce another term into the Einstein field equations, which will describe a non-gravitational field associated with the light-matter interaction. This term will play the same role as the cosmological constant $\Lambda$ in Eqs. (1-2), but in contrast to $\Lambda$, which is of unclear physical nature, the light-matter interaction term is physically well justified.

The light-matter interaction will be characterized by density $\rho_\Lambda$ and pressure $p_\Lambda$. The energy-momentum tensor $\Lambda^{\mu \nu}$ of the light-matter interaction will be defined in a similar way as the energy-momentum tensor $T^{\mu \nu}$ for the gravitational field, see Appendix A for details. Assuming that the Universe is filled by a perfect homogeneous and isotropic fluid and its expansion is described by the standard FLRW metric, we obtain the following modified Friedmann equations (see Eqs. A11 and A15 in Appendix A): 
\begin{equation}\label{eq17}
{\left({\frac{\dot a}{a}}\right)}^2 = \frac{8\pi G}{3} \rho - \frac{2}{3} \rho_\Lambda -\frac{k c^2}{a^2}\,, 
\end{equation}
\begin{equation}\label{eq18}
\frac{\ddot a}{a} =  -\frac{4 \pi G}{3} (\alpha-2) \rho + \frac{1}{3} (\beta-2) \rho_\Lambda \,.
\end{equation}
where coefficients $\alpha$ and $\beta$ define the dependence of densities $\rho$ and $\rho_\Lambda$ on the scale factor $a(t)$: $\rho \sim a^{-\alpha}$  and $\rho_\Lambda \sim a^{-\beta}$. Specifying Eq. (18) for the pressureless fluid $(\alpha = 3)$ and taking into account that $\rho_\Lambda = \kappa j/c$ , we obtain the final form of the Friedmann equations for the opaque universe
\begin{equation}\label{eq19}
{\left({\frac{\dot a}{a}}\right)}^2 = \frac{8\pi G}{3} \rho - \frac{2}{3} \frac{\kappa j}{c} -\frac{k c^2}{a^2}\,, 
\end{equation}
\begin{equation}\label{eq20}
\frac{\ddot{a}}{a} = -\frac{4\pi G}{3} \rho + \frac{\beta-2}{3} \frac{\kappa j}{c} \,. 
\end{equation}
Comparing Eqs (1-2) with Eqs (19-20), we see that the modified Friedmann equations can be rewritten into a form almost identical with the original Friedmann equations 
\begin{equation}\label{eq21}
{\left({\frac{\dot a}{a}}\right)}^2 = \frac{8\pi G}{3} \rho - \frac{k c^2}{a^2} + \frac{1}{3} \Lambda c^2 \,, 
\end{equation}
\begin{equation}\label{eq22}
\frac{\ddot{a}}{a} = -\frac{4\pi G}{3} \rho + \frac{2-\beta}{2} \frac{1}{3} \Lambda c^2 \,, 
\end{equation}
if the cosmological term $\Lambda$ is defined as
\begin{equation}\label{eq23}
\Lambda = 2 \frac{\kappa j}{c^3} \,. 
\end{equation}
The only difference is in factor $(2-\beta)/2$ in Eq. (22), originating from the fact that $\Lambda$ is not a constant any more but it depends on the scale factor $a(t)$. If $\beta = 0$, Eq. (22) becomes identical with the Friedmann equation (2).

\subsection{Distance-redshift relation}

Assuming that $\Lambda$ depends on $a$ as $\sim a^{-\beta}$ in Eq. (21), the Hubble parameter reads
\begin{equation}\label{eq24}
H^2\left(a\right) = H^2_0 \left[{\Omega_m a^{-3} + \Omega_r a^{-4}+ \Omega_a a^{-\beta} + \Omega_k a^{-2}}\right] \,,
\end{equation}
where $\Omega_m$, $\Omega_r$, $\Omega_a$ and $\Omega_k$ are the normalized matter, radiation, radiation-absorption and curvature terms, respectively. In contrast to $\Omega_m$ and $\Omega_r$, which describe attractive gravitational forces produced by matter and radiation in the Universe, $\Omega_a$ describes repulsive non-gravitational forces produced by the light-matter interaction. Since gravity associated with radiation is non-negligible only for $z > 1100$, we can assume $\Omega_r = 0$ and specify Eq. (24) for the matter-dominated opaque universe as
\begin{equation}\label{eq25}
H^2\left(a\right) = H^2_0 \left[{\Omega_m a^{-3} + \Omega_a a^{-\beta} + \Omega_k a^{-2}}\right] \,,
\end{equation}
with the condition
\begin{equation}\label{eq26}
\Omega_m + \Omega_a + \Omega_k = 1 \,, 
\end{equation}
where
\begin{equation}\label{eq27}
\Omega_m = \frac{1}{H^2_0} \left({ \frac{8 \pi G}{3} \rho_0}\right) \,,  
\end{equation}
\begin{equation}\label{eq28}
\Omega_a = -\frac{1}{H^2_0} \left(\frac{2}{3} \frac{\kappa_0 j_0}{c} \right) \,,
\end{equation}
\begin{equation}\label{eq29}
\Omega_k = -\frac{k c^2}{H^2_0} \,. 
\end{equation}
The minus sign in Eq. (28) means that the radiation pressure due to the light-matter interaction acts against the gravity. Considering $a = 1/(1+z)$, the comoving distance  is expressed from Eq. (25) as a function of redshift as follows
\begin{equation}\label{eq30}
\begin{split}
dr = \frac{c}{H_0} \frac{dz}{\sqrt{\Omega_m \left(1+z\right)^3 +  
\Omega_a \left(1+z\right)^\beta + \Omega_k \left(1+z\right)^2}} \,. 
\end{split}
\end{equation}
%

\subsection{Redshift dependence of the light-matter interaction}

The radiation-absorption term $\Lambda$ defined in Eq. (23) is redshift dependent. Under the assumption that the number of sources and their luminosity conserves in time, the rest-frame luminosity density $j_\nu$ for a given frequency $\nu$ depends on redshift as $(1+z)^3$ and the bolometric luminosity density $j$ depends on redshift as
\begin{equation}\label{eq31}
j = j_0 a^{-4} = j_0 \left(1 + z \right)^4 \,. 
\end{equation}
where subscript '0' corresponds to the quantity observed at present. The assumption of the independence of the global stellar mass in the Universe looks apparently unrealistic but it is fully consistent with observations if corrections to opacity of the high-redshift Universe are applied \citep{Vavrycuk2017a,Vavrycuk2018}, see Fig.~\ref{fig:3}. 

The luminosity density comprises energy radiated by galaxies into the intergalactic space and thermal radiation of intergalactic dust. All these sources produce cosmic background radiation in the Universe being the sum of the cosmic X-ray background (CXB), the extragalactic background light (EBL) and the cosmic microwave background (CMB). The cosmic background radiation as any radiation in the expanded universe depends on redshift as
\begin{equation}\label{eq32}
I = I_0 a^{-4} = I_0 \left(1 + z \right)^4 \,. 
\end{equation}

\begin{figure*}
\includegraphics[angle=0, width=16cm, trim=150 150 150 160, clip = true]{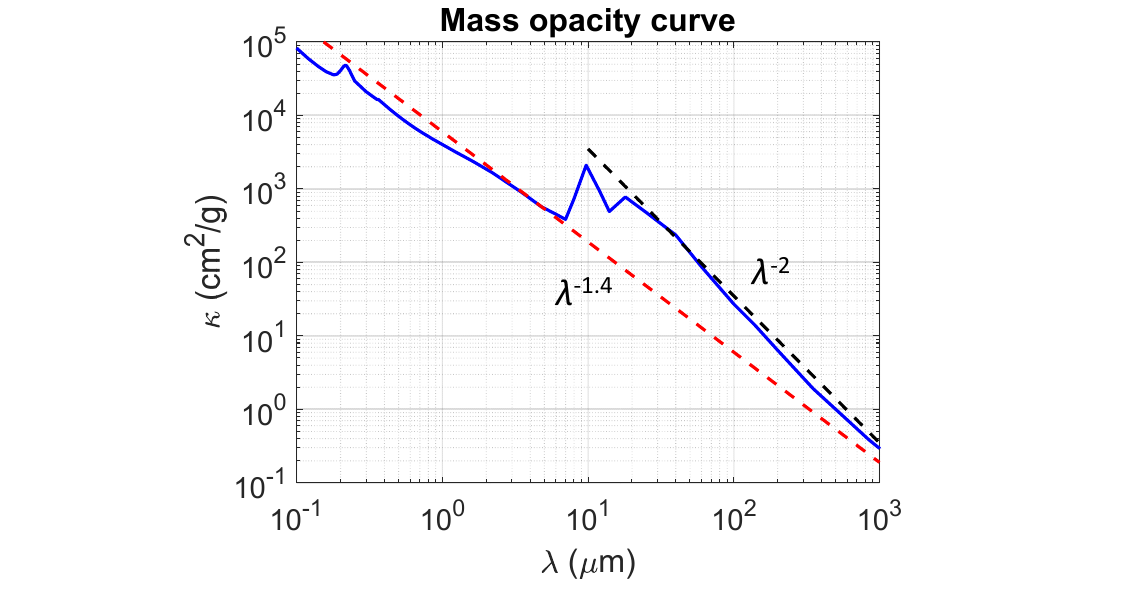}
\caption{
The mass opacity $\kappa$ as a function of wavelength for the so-called MRN dust model \citep{Mathis1977} defined by the power-law grains-size distribution with lower and upper size limits between $\sim5$ and $\sim250$ nm, see Tables 4-6 of \citet{Draine2003}. The black and red dashed lines show the long-wavelength asymptotic behaviour predicted by the power law with the spectral index of 2 and 1.4, respectively.
}
\label{fig:5}
\end{figure*}

Also the mass opacity $\kappa$ in Eq. (23) depends on redshift. Based on the extinction law, the mass opacity $\kappa$ depends on the wavelength $\lambda$ of absorbed radiation as $\lambda^{-\gamma}$, where $\gamma$ is the spectral index ranging between 1.0 and 2.0 for grains with size of 0.2 $\mu$m or smaller \citep{Mathis1977, Draine2011}, see Fig.~\ref{fig:5}. Hence, if radiation changes its wavelength due to the redshift, the opacity $\kappa$ is also redshift dependent. Consequently, the coefficient $\beta$ describing the redshift-dependent radiation-absorption term in Eqs (25) and (30) ranges from 5 to 6. By contrast, the mass opacity is wavelength independent for large grains with size larger than wavelength $\lambda$ and the radiation-absorption term depends on $z$ as $(1+z)^4$ only. 

%

Since the coefficient $\beta$ essentially affects behaviour of the Hubble parameter $H(a)$ and subsequently the evolution of the Universe, we will discuss the origins of its enormously high value in details. The normalized matter and radiation terms $\Omega_m$ and $\Omega_r$ in Eq. (24) depend on the scale factor $a$ as $a^{-3}$ and $a^{-4}$, respectively. Hence, one would intuitively expect that the interaction of matter with radiation will produce term $a^{-\beta}$ with $\beta$ ranging between 3 and 4. However, this speculation is false, because it ignores essential property of the radiation-matter interaction - its frequency dependence. The interaction of radiation with matter is caused by absorption of light by grains of cosmic dust, which depends on the wavelength of light and on the size of dust grains. While large wavelengths of light are absorbed weakly, the short wavelengths are absorbed more intensely. Hence, three effects are involved in the light-matter interaction: (1) an increase of the intensity of light as $(1+z)^{3}$ associated with decreasing the volume of the Universe with redshift, (2) an additional increase of the intensity of light as $(1+z)$ due to the shortening of wavelengths of photons caused by the cosmological redshift, and (3) an increase of light absorption as $(1+z)^{\gamma}$, with $\gamma$ ranging between 1 and 2, because the photons at high redshifts have shorter wavelengths and interact much strongly with cosmic dust grains than photons at the present epoch.

\subsection{Limits of the scale factor $a$}

In order to get simple closed-form formulas, we assume in the next that the mean spectral index $\gamma$ characterizing the absorption of light by mixture of grains of varying size is 1. Consequently, the radiation-absorption term depends on $a$ as $\sim a^{-5}$. The scale factor $a$ of the Universe with the zero expansion rate is defined by the zero Hubble parameter in Eq. (25), which yields a cubic equation in $a$
\begin{equation}\label{eq33}
\Omega_k a^3 + \Omega_m a^2 + \Omega_a = 0 \,. 
\end{equation}
Taking into account that $\Omega_m > 0$ and $\Omega_a < 0$, Eq. (33) has two distinct real positive roots for 
\begin{equation}\label{eq34}
\left({\frac{\Omega_m}{3}}\right)^2 > \left({\frac{\Omega_k}{2}}\right)^2 \left|\Omega_a\right| \,\,\, 
\mathrm{and} \,\,\, 
\Omega_k < 0 \,. 
\end{equation}
Negative $\Omega_a$ and $\Omega_k$ imply that
\begin{equation}\label{eq35}
\Omega_m > 1 \,\,\,
\mathrm{and} \,\,\, 
\rho_0 > \rho_c = \frac{8 \pi G}{3 H^2_0} \,. 
\end{equation}
Under these conditions, Eq. (25) describes a universe with a cyclic expansion/contraction history and the two real positive roots $a_\mathrm{min}$ and $a_\mathrm{max}$ define the minimum and maximum scale factors of the Universe. For $\Omega_a \ll 1$, the scale factors $a_\mathrm{min}$ and $a_\mathrm{max}$ read approximately
\begin{equation}\label{eq36}
a_\mathrm{min} \cong \sqrt{\left|{\frac{\Omega_a}{\Omega_m}}\right|} \,\,\, \mathrm{and} \,\,\, 
a_\mathrm{max} \cong       \left|{\frac{\Omega_m}{\Omega_k}}\right| \,, 
\end{equation}
and the maximum redshift is
\begin{equation}\label{eq37}
z_\mathrm{max} = \frac{1}{a_\mathrm{min}} - 1 \,. 
\end{equation}
The scale factors $a$ of the Universe with the maximum expansion/contraction rates are defined by
\begin{equation}\label{eq38}
\frac{d}{da} H^2 \left(a\right) = 0 \,, 
\end{equation}
which yields a cubic equation in $a$
\begin{equation}\label{eq39}
2 \Omega_k a^3 + 3 \Omega_m a^2 + 5 \Omega_a = 0 \,. 
\end{equation}
Taking into account Eqs (21-22) and Eqs (27-29), the deceleration of the expansion reads
\begin{equation}\label{eq40}
\ddot a = -\frac{1}{2} H^2_0 \left[\Omega_m a^{-2} + 3 \Omega_a a^{-4}\right] \,. 
\end{equation}
Hence, the zero deceleration is for the scale factor
\begin{equation}\label{eq41}
a = \sqrt{\left|{\frac{3\Omega_a}{\Omega_m}}\right|} \,. 
\end{equation}

The above equations are quite simple, because they are derived for the spectral index $\gamma =1$. For other values of $\gamma$, the limits of the scale factor $a$ are obtained by solving the equation for the zero Hubble parameter numerically. In general, the higher spectral index $\gamma$, the smaller value of the maximum redshift $z_{\mathrm{max}}$, see the next sections.

\section{Parameters for modelling}

For calculating the expansion history and cosmic dynamics of the Universe, we need observations of the mass opacity of intergalactic dust grains, the galaxy luminosity density, the mean mass density, and the expansion rate and curvature of the Universe at the present time.

\subsection{Mass opacity of cosmic dust}

When estimating the mass opacity of dust, $\kappa_0$, we have to know basic parameters of dust grains. The size $d$ of dust grains is in the range of $0.01 - 0.2 \, \mu$m with a power-law distribution $d^{-q}$ with $q = 3.5$ \citep{Mathis1977, Jones1996}, but silicate and carbonaceous grains dominating the scattering are typically with $d \approx 0.1 \, \mu$m \citep{Draine_Fraisse2009, Draine2011}. The grains of size  $0.07 \, \mu\mathrm{m} \leq d \leq 0.2 \, \mu\mathrm{m}$ are also ejected to the IGM most effectively \citep{Davies1998, Bianchi_Ferrara2005}. The grains form complicate fluffy aggregates, which are often elongated or needle-shaped \citep{Wright1982, Wright1987}. Considering the density of carbonaceous material $\rho \approx 2.2 \, \mathrm{g \, cm}^{-3}$ and the silicate density $\rho \approx 3.8 \, \mathrm{g \, cm}^{-3}$ \citep{Draine2011}, the average density of porous dust grains is $\approx 2 \, \mathrm{g \, cm}^{-3}$ or less \citep{Flynn1994, Kocifaj1999, Kohout2014}. Consequently, the standard dust models \citep{Weingartner_Draine2001} predict the wavelength-dependent mass opacity. For example, \citet{Draine2003} reports the mass opacity of 855 $\mathrm {m}^2\mathrm{kg}^{-1}$ at the V-band and the mass opacity of 402 $\mathrm {m}^2\mathrm{kg}^{-1}$ for a wavelength of 1 $\mu$m, which corresponds to the maximum intensity of the EBL.

\subsection{EBL and the galaxy luminosity density}
The EBL covers a wide range of wavelengths from 0.1 to 1000 $\mu$m. It was measured, for example, by the IRAS, FIRAS, DIRBE on COBE, and SCUBA instruments; for reviews, see \citet{Hauser2001, Lagache2005, Cooray2016}. The direct measurements are supplemented by integrating light from extragalactic source counts \citep{Madau2000, Hauser2001} and by attenuation of gamma rays from distant blazars due to scattering on the EBL \citep{Kneiske2004, Dwek2005, Primack2011, Gilmore2012}. The EBL spectrum has two maxima: associated with the radiation of stars (at $0.7 - 2 \, \mu$m) and with the thermal radiation of dust in galaxies (at $100 - 200 \, \mu$m), see \citet{Schlegel1998, Calzetti2000}. Despite extensive measurements, uncertainties of the EBL are still large. The total EBL should fall between 40 and 200 $\mathrm{n W\, m^{-2}\,sr^{-1}}$ \citep[his fig. 1]{Vavrycuk2018} with the most likely value $I^\mathrm{EBL} = 80 - 100 \,\, \mathrm{nW\,m^{-2}\,sr^{-1}}$ \citep{Hauser2001, Bernstein2002a, Bernstein2002b, Bernstein2002c, Bernstein2007}. 

The galaxy luminosity density is determined from the Schechter function \citep{Schechter1976}. It has been measured by large surveys 2dFGRS \citep{Cross2001}, SDSS \citep{Blanton2001, Blanton2003} or CS \citep{Brown2001}. The luminosity function in the R-band was estimated at $z = 0$ to be $\left(1.84 \pm 0.04\right) \times 10^8 \,\,h \, L_\odot \,\, \mathrm{Mpc^{-3}}$ for the SDSS data \citep{Blanton2003} and $\left(1.9 \pm 0.6\right) \times 10^8 \,\, h \, L_\odot \,\, \mathrm{Mpc^{-3}}$ for the CS data \citep{Brown2001}. The bolometric luminosity density is estimated by considering the spectral energy distribution (SED) of galaxies averaged over different galaxy types, being thus about 1.7 times larger than that in the R-band \citep[his table 2]{Vavrycuk2017a}:
$j_0 \approx 3.1 \times 10^8 \,\, h\, L_\odot \, \mathrm{Mpc^{-3}}$.

\subsection{Matter density of the Universe}

A simplest and most straightforward method, how to estimate the matter density, is based on galaxy surveys and computation of the mass from the observed galaxy luminosity and from the mass-to-light ratio ($M/L$) that reflects the total amount of the mass relative to the light within a given scale. The $M/L$ ratio is, however, scale dependent and increases from bright, luminous parts of galaxies to their halos (with radius of $\approx 200$ kpc) formed by (baryonic and/or speculative non-baryonic) dark matter. The $M/L$ ratio depends also on a galaxy type being about 3 to 4 times larger for elliptical/SO galaxies than for typical spirals, hence the observed $M/L_B$ is $\approx 100 \, h$ for spirals, but $\approx 400 \, h$ for ellipticals at radius of $\approx 200$ kpc, see \citet{Bahcall1995}. Considering the mean asymptotic ratio $M/L_B$ of $200 - 300 \, h$ and the observed mean luminosity density of the Universe at $z = 0$ of $\approx 2.5 \times 10^8 \,\, h \, L_\odot \, \mathrm{Mpc^{-3}}$ reported by \citet{Cross2001}, the matter density $\Omega_m$ associated with galaxies is about $0.2 - 0.3$ ($\Omega_m = 1$ means the critical density).

Another source of matter in the universe is connected to Ly$\alpha$ absorbers containing photoionized hydrogen at $\approx 10^4$ K and being detected by the Ly$\alpha$ forest in quasar spectra \citep{Meiksin2009}. These systems are partly located in the galaxy halos, but a significant portion of them cannot be associated to any galaxy, being observed, for example, in voids \citep{Penton2002, Tejos2012, Tejos2014}. The Ly$\alpha$ absorbers also form the intragroup and intracluster medium \citep{Bielby2017} and the intergalactic medium nearby the other large-scale galaxy structures like the galaxy filaments \citep{Tejos2014, Wakker2015}. In addition, it is speculated that large amount of matter is located in the warm-hot intergalactic medium (WHIM) that is a gaseous phase of moderate to low density ($\approx 10 - 30$ times the mean density of the Universe) and at temperatures of $10^5 - 10^7$ K. Although it is difficult to observe the WHIM because of low column densities of HI in the hot gas, they might be potentially detected by surveys of broad HI Ly$\alpha$ absorbers (BLAs) as reported by \citet{Nicastro2018} or \citet{Pessa2018}. 

Hence, we conclude that the estimate of matter density $\Omega_m = 0.2 - 0.3$ inferred from a distribution of galaxies is just a lower limit, while the upper limit of $\Omega_m$ is unconstrained being possibly close to or even higher than 1. This statement contradicts the commonly accepted value of $\Omega_m = 0.3$ reported by \citet{ Ade2015_Planck_XIII_Cosmological_parameters, Aghanim2018_Planck_VI_Cosmological_parameters} which is based on the interpretation of the CMB as a relic radiation of the Big Bang. 

\subsection{Hubble constant and cosmic curvature}

The Hubble constant $H_0$ is measured by methods based on the Sunyaev-Zel'dovich effect \citep{Birkinshaw1999, Carlstrom2002, Bonamente2006} or gravitational lensing \citep{Suyu2013, Bonvin2017}, gravitational waves \citep{Abbott2017, Vitale_Chen2018, Howlett_Davis2020} or acoustic peaks in the CMB spectrum provided by \citet{Ade2015_Planck_XIII_Cosmological_parameters}, and they yield values mostly ranging between 67 and 74 $\mathrm{km \, s^{-1} \, Mpc^{-1}}$. Among these approaches, direct methods are considered to be most reliable and accurate (for a review, see \citet{Jackson2015}). These methods are based on measuring local distances up to $20 - 30$ Mpc using Cepheid variables observed by the Hubble Space Telescope (HST). The HST galaxies with distance measured with the Cepheid variables are then used to calibrate the SNe Ia data. With this calibration, the distance measure can be extended to other more distant galaxies (hundreds of Mpc) in which SNe Ia are detected \citep{Freedman2001, Riess2011}. The estimate of $H_0$ obtained by \citet{Riess2016} using the Cepheid calibration is $73.25 \pm 1.74 \,\, \mathrm{km \, s^{-1} \, Mpc^{-1}}$. The precision of the distance scale was further reduced by a factor of 2.5 by \citet{Riess2018}. Another estimate of $H_0$ obtained by \citet{Freedman2019} using the SNe Ia with a red giant branch calibration is $69.8 \pm 2.5 \,\, \mathrm{km \, s^{-1} \, Mpc^{-1}}$. 

Assuming the $\Lambda$CDM model, the CMB and BAO observations indicate a nearly flat Universe \citep{Ade2015_Planck_XIII_Cosmological_parameters}. This method is not, however, model independent and ignores an impact of cosmic dust on the CMB. A model-independent method proposed by \citet{Clarkson2007} is based on reconstructing the comoving distances by Hubble parameter data and comparing with the luminosity distances \citep{Li2016, Wei_Wu2017} or the angular diameter distances \citep{Yu_Wang2016}. The cosmic curvature can also be constrained using strongly gravitational lensed SNe Ia \citep{Qi2019} and using lensing time delays and gravitational waves \citep{Liao2019}. The authors report the curvature term $\Omega_k$ ranging between -0.3 to 0 indicating a closed universe, not significantly departing from flat geometry.

\begin{figure*}
\includegraphics[angle=0,width=12.0cm, clip=true, trim=80 50 120 50]{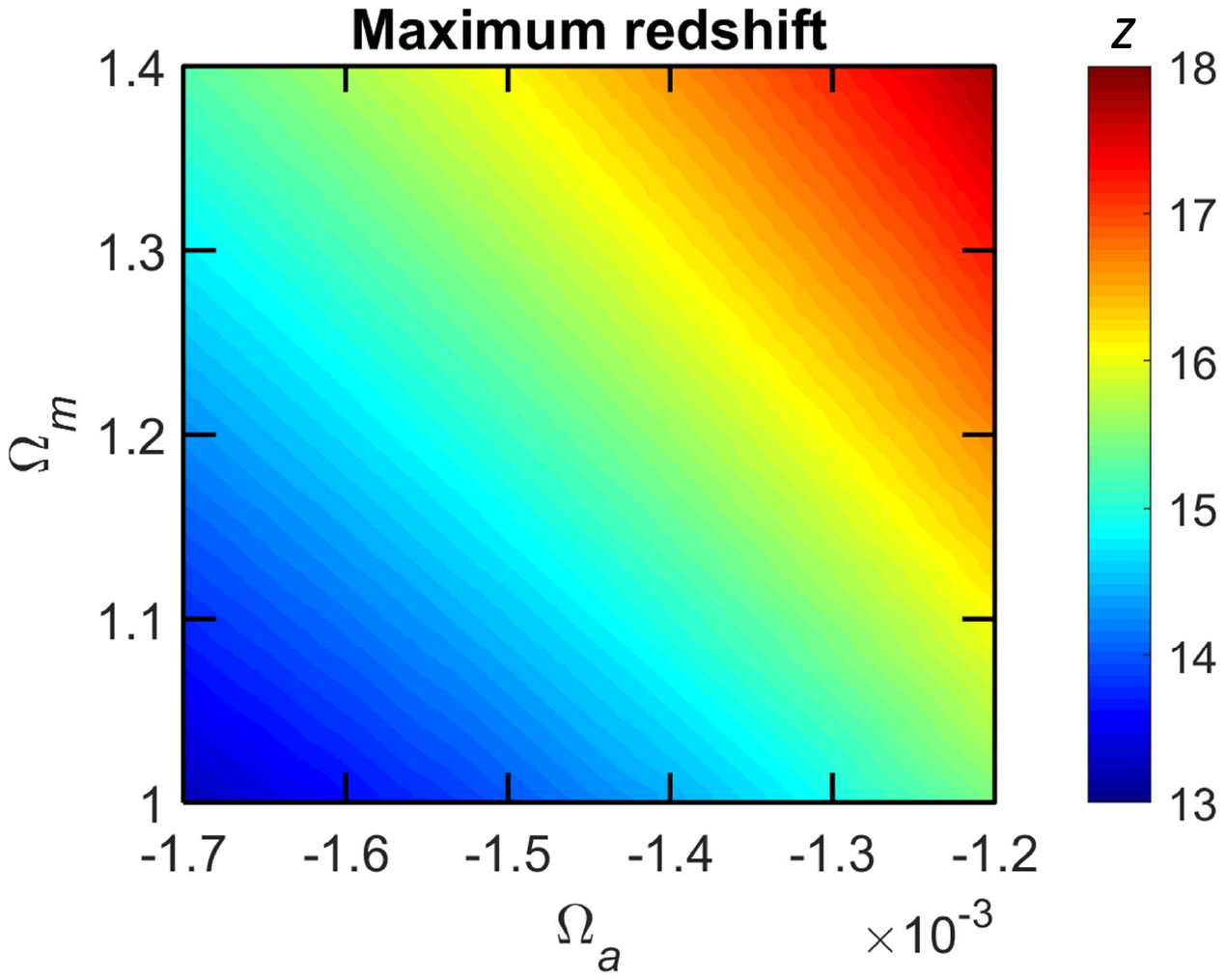}
\caption{
Maximum redshift as a function of $\Omega_m$ and $\Omega_a$. The power-law exponent $\beta$ describing a decay of the radiation-absorption term with the scale factor $a$ is assumed to be 5.4, see Table 1.
}
\label{fig:6}
\end{figure*}

\section{Results}

\begin{table*}
%
%
\caption[]{Maximum redshift and scale factor in the cyclic model of the opaque universe}  

\label{Table:1}      
%
\begin{tabular}{|c| c c c c c |c c|}  
%
%
\hline
%
%
& \multicolumn{5}{c|}{Input parameters} & \multicolumn{2}{c|}{Output} \\
\cline{2-8}
 Model & $\varepsilon$ & $\Omega_m$ & $\Omega_a$ & $\beta$ & $\Omega_k$  & $a_\mathrm{max}$ & $z_\mathrm{max}$\\
 & & &  &  &  &  &  \\
%
%
\hline                        
%
%
A & 6  &  1.2 & $\,\, -1.7 \times 10^{-3}$ & 5.6 & $\,\, -0.198$ &  6.1 & 11.4 \\
B & 4  &  1.2 & $\,\, -1.2 \times 10^{-3}$ & 5.2 & $\,\, -0.199$ &  6.0 & 22.0 \\    
C & 5  &  1.2 & $\,\, -1.5 \times 10^{-3}$ & 5.4 & $\,\, -0.199$ &  6.0 & 15.1 \\
D & 5  &  1.1 & $\,\, -1.5 \times 10^{-3}$ & 5.4 & $\,\, -0.099$ & 11.2 & 14.6 \\
E & 5  &  1.3 & $\,\, -1.5 \times 10^{-3}$ & 5.4 & $\,\, -0.299$ &  4.4 & 15.6 \\ 
%
%
\hline                                  
\end{tabular}
%
%
\\
\begin{tablenotes}
\item{Parameter $\varepsilon$ is the ratio of the spheroidal to spherical dust grain cross-sections, $\Omega_m$, $\Omega_a$, and $\Omega_k$ are the matter, radiation-absorption and curvature terms, $\beta$ is the power-law exponent describing a decay of the radiation-absorption term with the scale factor $a$ in Eq. (25), and $a_\mathrm{max}$ and $z_\mathrm{max}$ are the estimates of the maximum scale factor and redshift, respectively. Models A, B and C predict low, high and optimum values of $z_\mathrm{max}$. Models E, D and C predict low, high and optimum values of $a_\mathrm{max}$.} 
\end{tablenotes} 
\end{table*}
%
\begin{figure*}
\includegraphics[angle=0, width=16cm, trim=50 170 80 120, clip = true]{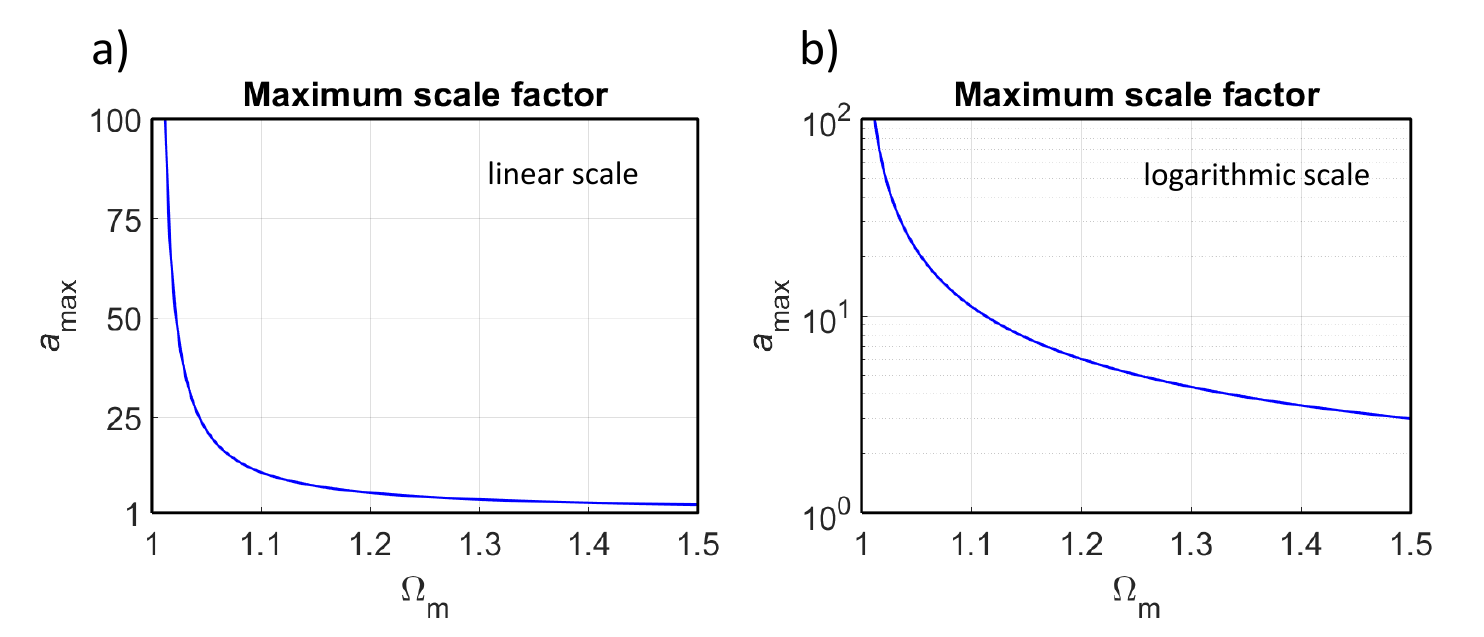}
\caption{
The maximum scale factor as a function of $\Omega_m$. (a) Linear scale, (b) logarithmic scale. The dependence on $\Omega_a$ is negligible.
}
\label{fig:7}
\end{figure*}

\begin{figure*}
\includegraphics[angle=0,width=16 cm, clip=true, trim=60 120 60 160]{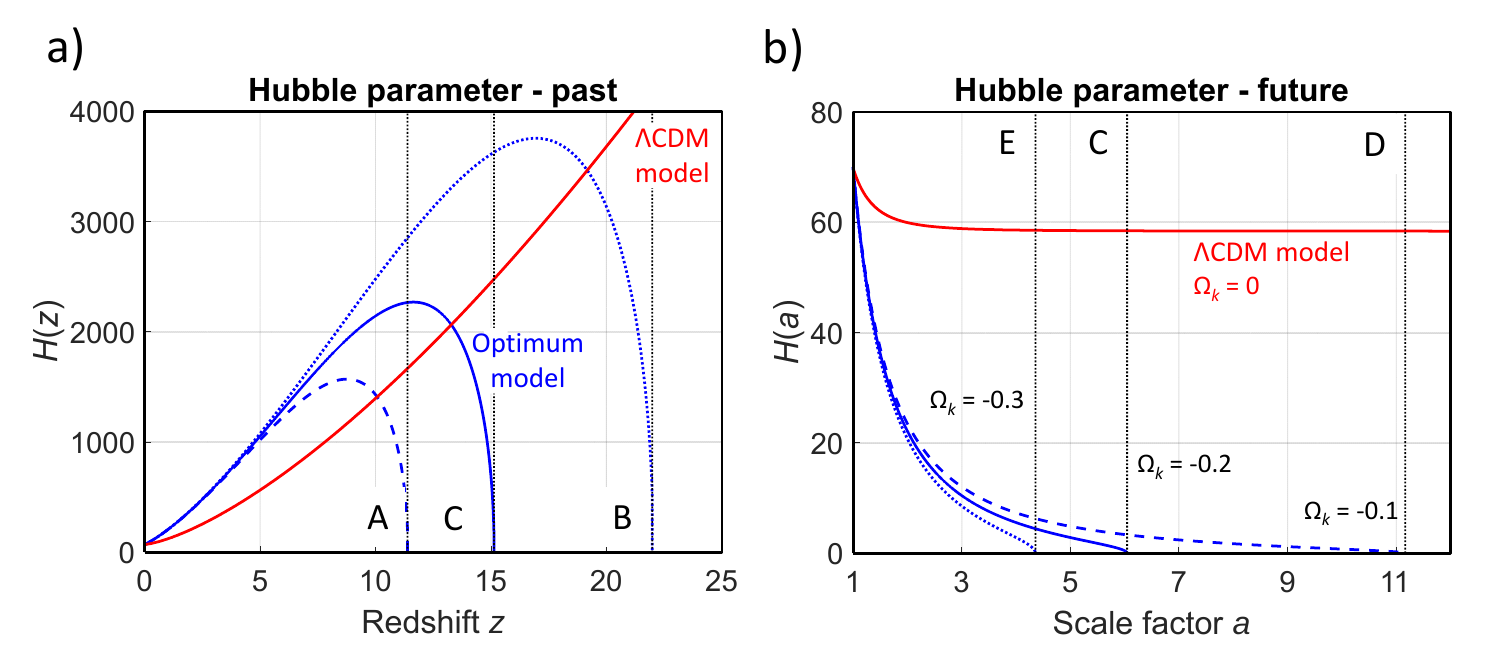}
\caption{
The evolution of the Hubble parameter with redshift in the past and with the scale factor in the future (in $\mathrm{km \, s^{-1} \, Mpc^{-1}}$). (a) The blue dashed, dotted and solid lines show Models A, B and C in Table 1. (b) The blue solid, dashed, and dotted lines show Models C, D and E in Table 1. The black dotted lines mark the predicted maximum redshifts (a) and maximum scale factors (b) for the models considered. The red solid line shows the flat $\Lambda$CDM model with $H_0 = 69.8 \,\, \mathrm{km \, s^{-1} \, Mpc^{-1}}$, taken from \citet{Freedman2019}, and with $\Omega_m = 0.3$ and $\Omega_\Lambda = 0.7$.
}
\label{fig:8}
\end{figure*}

\begin{figure*}
\includegraphics[angle=0, width=12.0cm, trim= 10 60 180 100]{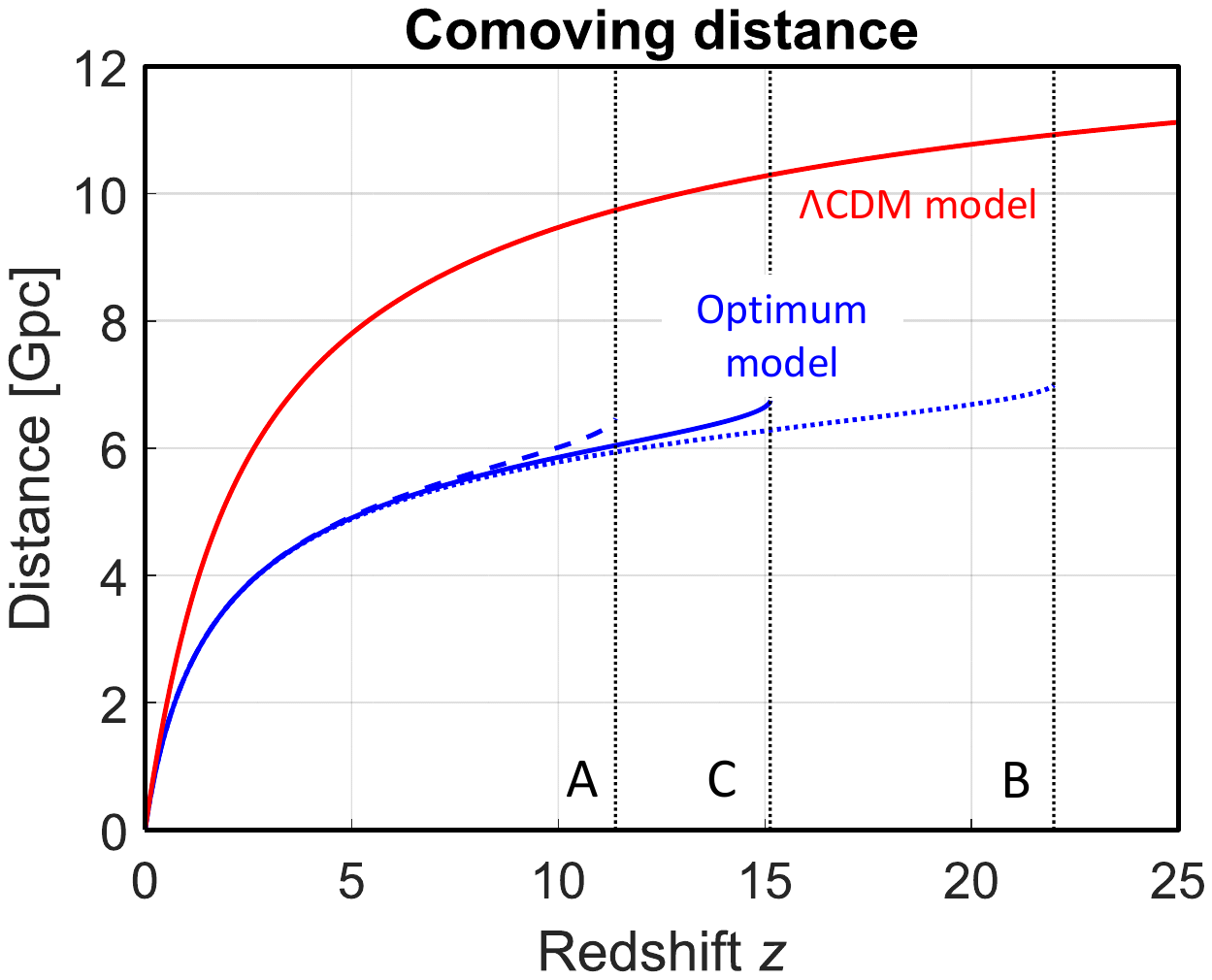}
\caption{
Comoving distance as a function of redshift $z$. The blue dashed, dotted and solid lines show Models A, B and C in Table 1. The black dotted lines mark the predicted maximum redshifts for the models considered. The red solid line shows the flat $\Lambda$CDM model with $H_0 = 69.8 \,\, \mathrm{km \, s^{-1} \, Mpc^{-1}}$, taken from \citet{Freedman2019}, and with $\Omega_m = 0.3$ and $\Omega_\Lambda = 0.7$.
}
\label{fig:9}
\end{figure*}

Estimating the required cosmological parameters from observations, the upper and lower limits of the volume of the Universe and the evolution of the Hubble parameter with time can be calculated using Eqs (25-29). The mass density of the Universe higher than the critical density is considered, and subsequently $\Omega_m$ is higher than 1. The Hubble constant is $H_0 = 69.8 \, \mathrm{km \, s^{-1} \, Mpc^{-1}}$, taken from \citet{Freedman2019}. The mass opacity $\kappa_0$ of 402 $\mathrm{m}^2 \mathrm{kg}^{-1}$ is taken from table 4 of \citet{Draine2003} and it characterizes the opacity of dust at a wavelength of 1 $\mu$m. The opacity is further multiplied by factor $\varepsilon$ reflecting that dust grains are not spherical but rather prolate spheroids having a larger effective cross-section. The luminosity density is $j_0 = 3.1 \times 10^8 \,\, h \, L_\odot \,\, \mathrm{Mpc^{-3}}$. The radiation-absorption term in Eq. (28) is multiplied by a factor of 2, because photons are not only absorbed but also radiated by dust grains to maintain the thermal equilibrium. The exponent $\beta$ of the power-law decay of the radiation-absorption term in Eq. (25) ranges from 5.2 to 5.6. The results of modelling are summarized in Table 1.

As seen in Fig.~\ref{fig:6}, the maximum redshift of the Universe depends on $\Omega_m$ and $\Omega_a$, and ranges from 13 to 18 for $\beta = 5.4$. In contrast to $a_\mathrm{min}$ depending on both  $\Omega_m$ and $\Omega_a$, the maximum scale factor $a_\mathrm{max}$ of the Universe depends primarily on $\Omega_m$ only. Fig.~\ref{fig:7} shows that $a_\mathrm{max}$ rapidly decreases with increasing $\Omega_m$. Obviously, the limiting value is $\Omega_m = 1$, when $a_\mathrm{max}$ is infinite. The limiting value is $\Omega_m = 1$, when $a_\mathrm{max}$ is infinite. For $\Omega_m = 1.1$, 1.2, 1.3 and 1.5, the scale factor $a_\mathrm{max}$ is 11.2, 6.0, 4.4 and 3.0, respectively.

The history of the Hubble parameter $H(z)$ and its evolution in the future $H(a)$ calculated by Eq. (25) is shown in Fig.~\ref{fig:8} for five scenarios summarized in Table 1. The form of $H(z)$ in Fig.~\ref{fig:8}a is controlled by $\Omega_a$ and the power-law exponent $\beta$, while the form of $H(a)$ in Fig.~\ref{fig:8}b is controlled by $\Omega_m$. The Hubble parameter $H(z)$ increases with redshift up to its maximum. After that the function rapidly decreases to zero. The drop of $H(z)$ is due to a fast increase of light attenuation producing strong repulsive forces at high redshift. For future epochs, function $H(a)$ is predicted to monotonously decrease to zero. The rate of decrease is controlled just by gravitational forces; the repulsive forces originating in light attenuation are negligible. For a comparison, Fig.~\ref{fig:8} (red line) shows the Hubble parameter  $H(a)$ for the standard $\Lambda$CDM model  \citep{Ade2015_Planck_XIII_Cosmological_parameters}, which is described by Eq. (6) with $\Omega_m = 0.3$ and $\Omega_\Lambda = 0.7$.  

The distance-redshift relation of the proposed cyclic model of the Universe is quite different from the standard $\Lambda$CDM model (see Fig.~\ref{fig:9}). In both models, the comoving distance monotonously increases with redshift, but the redshift can go possibly to 1000 or more in the standard model, while the maximum redshift is likely 14-15 in the optimum cyclic model. The increase of distance with redshift is remarkably steeper for the $\Lambda$CDM model than for the cyclic model. The ratio between distances in the cyclic and $\Lambda$CDM models rapidly decreases from 1 at $z = 0$ to about 0.63 at $z > 4$.

\section{Other supporting evidence}
The cyclic cosmological model of the opaque universe successfully removes some tensions of the standard $\Lambda$CDM model: 

\begin{itemize}
\item
{The model does not limit the age of stars in the Universe. For example, observations of a nearby star HD 140283 \citep{Bond2013} with age of $14.46 \pm 0.31 \, \mathrm{Gyr}$ are in conflict with the age of the Universe, $13.80 \pm 0.02 \, \mathrm{Gyr}$, determined from the interpretation of the CMB as relic radiation of the Big Bang \citep{Ade2015_Planck_XIII_Cosmological_parameters}.}

\item
{The model predicts the existence of very old mature galaxies at high redshifts. The existence of mature galaxies in the early Universe was confirmed, for example, by \citet{Watson2015} who analyzed observations of the Atacama Large Millimetre Array (ALMA) and revealed a galaxy at $z > 7$ highly evolved with a large stellar mass and heavily enriched in dust. Similarly, \citet{Laporte2017} analyzed a galaxy at $z \approx 8$ with a stellar mass of $\approx 2 \times 10^9 \,\, M_\odot$ and a dust mass of $\approx 6 \times 10^6 \,\, M_\odot$. Large amount of dust is reported by \citet{Venemans2017} for a quasar at $z = 7.5$ in the interstellar medium of its host galaxy. In addition, a remarkably bright galaxy at $z \approx 11$ was found by \citet{Oesch2016} and a significant increase in the number of galaxies for $8.5 < z < 12$ was reported by \citet{Ellis2013}. Note that the number of papers reporting discoveries of galaxies at $z \approx 10$ or higher is growing rapidly \citep{Hashimoto2018, Hoag2018, Oesch2018, Salmon2018}.}

\item
{The model is capable to explain the SNe Ia dimming discovered by \citet{Riess1998} and \citet{Perlmutter1999} without introducing dark energy as the hypothetical energy of vacuum \citep{Vavrycuk2019}, which is difficult to explain under quantum field theory \citep{Weinberg2013}. Moreover, the speed of gravitational waves and the speed of light differ for most of dark energy models \citep{Sakstein_Jain2017, Ezquiaga_Zumalacarregui2017}, but observations of the binary neutron star merger GW170817 and its electromagnetic counterparts proved that both speeds coincide with a high accuracy.}

\item
{The model avoids a puzzle, how the CMB as relic radiation could survive the whole history of the Universe without any distortion \citep{Vavrycuk2017b} and why several unexpected features at large angular scales such as non-Gaussianity \citep{Vielva2004, Cruz2005, Ade2014_Planck_XXIV_Primordial} and a violation of statistical isotropy and scale invariance are observed in the CMB.}

\item
{The temperature of the CMB as thermal radiation of cosmic dust is predicted with the accuracy of 2\%, see \citet{Vavrycuk2018}. The CMB temperature is controlled by the EBL intensity and by the ratio of galactic and intergalactic opacities. The temperature of intergalactic dust increases linearly with redshift and exactly compensates the change of wavelengths due to redshift. Consequently, dust radiation looks apparently like the blackbody radiation with a single temperature.}

\item
{The model explains satisfactorily: (1) the observed bolometric intensity of the EBL with a value of $\approx100 \,\, \mathrm{nW \, m^{-1} \, sr^{-1}}$, see \citet{Vavrycuk2017a}, (2) the redshift evolution of the comoving UV luminosity density with extremely high values at redshifts $2 < z < 4$, see \citet{Vavrycuk2018}(his fig. 11), and (3) a strong decay of the global stellar mass density at high redshifts, see \citet{Vavrycuk2018} (his fig. 12). The increase of the luminosity density at $z \approx 2 - 3$ does not originate in the evolution of the star formation rate as commonly assumed but in the change of the proper volume of the Universe. The decrease of the luminosity density at high $z$ originates in the opacity of the high-redshift universe.}  

\end{itemize}

Note that the prediction of a close connection between the CMB anisotropies and the large-scale structures is common to both the standard model and the opaque universe model. The arguments are, however, reversed. The Big Bang theory assumes that the large-scale structures are a consequence of the CMB fluctuations originated at redshifts $z \approx 1100$, while the opaque universe model considers the CMB fluctuations as a consequence of the large-scale structures at redshifts less than $3-5$. The polarization anomalies of the CMB correlated with temperature anisotropies are caused by the polarized thermal emission of needle-shaped conducting dust grains aligned by large-scale magnetic fields around clusters and voids. The phenomenon is analogous to the polarized interstellar dust emission in our Galaxy, which is observed at shorter wavelengths because the temperature of the galactic dust is higher than that of the intergalactic dust  \citep{Lazarian2002, Gold2011, Ichiki2014, Ade2015_Planck_XIX_Polarized_emission, Aghanim2016_Planck_XLVIII_Dust_emission}.

\section{Discussion}
The standard Friedmann equations were derived for the transparent universe and assume no light-matter interaction. The equations contain densities $\Omega_m$ and $\Omega_r$ that describe the effects of gravity produced by matter, radiation and radiation pressure of photon gas. Since radiation pressure represents energy, it produces also gravity according to general relativity. The effects of radiation are, however, significant only for $z > 1100$. The modified Friedmann equations contain another density $\Omega_a$, which is also connected with the radiation pressure but in a different way. This pressure is produced by absorption of photons by ambient cosmic dust and it acts against gravity.

The radiation pressure as a cosmological force acting against the gravity has not been proposed yet, even though its role is well known in the stellar dynamics  \citep{Kippenhahn2012}. The radiation pressure is important in the evolution of massive stars \citep{Zinnecker_Yorke2007}, in supernovae stellar winds and in galactic wind dynamics \citep{Aguirre1999b, Martin2005, Hopkins2012, Hirashita_Inoue2019}. Apparently, the radiation pressure in the evolution of the Universe was overlooked, because the Universe was assumed to be transparent. By contrast, the role of radiation pressure is essential in the opaque universe model, because it is produced by absorption of photons by cosmic dust. Since the cosmic opacity and the intensity of the EBL steeply rise with redshift (see Fig. 2), the radiation pressure, negligible at present, becomes significant at high redshifts and can fully eliminate gravity and stop the universe contraction. In this process, small dust grains will probably be more important, because the mass opacity responsible for the radiation pressure rapidly increases with decreasing size of grains. Similarly, the emission of high-energy photons will affect the universe dynamics more distinctly than the photons re-emitted by dust grains which form the CMB. The high-energy photons emitted by stars are absorbed by 3-4 orders
more efficiently compared to the CMB photons, which are absorbed by dust very weakly.

Hence, the expansion/contraction evolution of the Universe might be a result of imbalance of gravitational forces and radiation pressure. Since the comoving global stellar and dust masses are basically independent of time with minor fluctuations only (see Fig. 3), the evolution of the Universe is stationary. The age of the Universe in the cyclic model is unconstrained and galaxies can be observed at any redshift less than the maximum redshift $z_\mathrm{max}$. The only limitation is high cosmic opacity, which can prevent observations of the most distant galaxies. Hypothetically, it is possible to observe galaxies from the previous cycle/cycles, if their distance is higher than that corresponding to $z_\mathrm{max} \approx 14-15$. The identification of galaxies from the previous cycles will be, however, difficult, because their redshift will be a periodic function with increasing distance. 

Obviously, a role of recycling processes is much more important in the cyclic cosmological model than in the Big Bang theory. The processes of formation/destruction of galaxies and their interaction with the circumgalactic medium through galactic winds and outflows \citep{Muzahid2015, Hatch2016, Segers2016, Angles-Alcazar2017, Muratov2017, Brennan2018} should play a central role in this model. Similarly, the formation of metals in nuclear fusion should be balanced in the long term by their destruction invoked, for example, by quasars. Indications supporting that such a scenario is not ruled out are provided by studies of metallicity with cosmic time, when observations do not show convincing evidence of the metallicity evolution. By contrast, they indicate \citep{Rauch1998, Pettini2004, Meiksin2009} a widespread metal pollution of the intergalactic medium in all epochs of the Universe and a failure to detect a pristine material with no metals at high redshifts.

In summary, the opaque universe model and the Big Bang theory are completely different concepts of the Universe. Both theories successfully predict basic astronomical observations such as the universe expansion, the luminosity density evolution with redshift, the global stellar mass history, the SNe Ia measurements and the CMB observations. However, the Big Bang theory needs the existence of dark matter and dark energy, which are supported by no firm evidence. Moreover, they contradict small-scale observations in galaxies  \citep{Kroupa2012, Kroupa2015, Buchert2016, Bullock_Boylan-Kolchin2017} and are disfavoured by observations of gravitational waves \citep{Ezquiaga_Zumalacarregui2017}. By contrast, the model of the eternal cyclic universe with high-redshift opacity is based on the standard physics, it is less speculative and predicts the current observations comparably well with no free parameters such as dark energy or dark matter. Nevertheless, this model opens other fundamental questions, such as about recycling processes of stars, galaxies and other objects in the Universe or about similarity/dissimilarity of individual cycles.

\section*{Acknowledgements}
I thank very much for constructive and fruitful comments and suggestions of the Editor and three anonymous referees.

\appendix
\section{Considering light-matter interactions in Einstein equations of general relativity}

The Einstein field equations read
\begin{equation}\label{eqA1}
G^{\mu\nu} + \Lambda g^{\mu\nu} = \frac{8 \pi G}{c^4} T^{\mu\nu} \,. 
\end{equation}
where $G^{\mu\nu}$ is the Einstein tensor, $\Lambda$ is the cosmological constant, $g^{\mu\nu}$ is the metric tensor, $G$ is the gravitational constant, $c$ is the speed of light, and $T^{\mu\nu}$ is the energy-momentum tensor. The Einstein tensor $G^{\mu\nu}$ describes the curvature of the spacetime associated with gravity produced by the presence of matter and/or energy described by the  energy-momentum tensor $T^{\mu\nu}$. 

Since ${G^{\mu\nu}}_{;\nu} = 0$ and ${g^{\mu\nu}}_{;\nu} = 0$, we get
\begin{equation}\label{eqA2}
{T^{\mu\nu}}_{;\nu} = 0 \,, 
\end{equation}
which expresses the energy-momentum conservation law. 

The cosmological constant $\Lambda$ in Eq. (A1) represents a non-gravitational field acting against the gravity. It was inserted into the equations by \citet{Einstein1917} in order to maintain the Universe to be static. Since the physical nature of $\Lambda$ was unclear, Einstein assumed the cosmological term in the simplest possible form. However, other forms of the cosmological term are, in principle, admissible. Obviously, the validity of the field equations (A1) is kept, if the cosmological term $\Lambda g^{\mu\nu}$ is substituted by the following more general term $\psi \Lambda^{\mu\nu}$, 
\begin{equation}\label{eqA3}
G^{\mu\nu} + \psi \Lambda^{\mu\nu} = \frac{8 \pi G}{c^4} T^{\mu\nu} \,, 
\end{equation}
where $\psi$ is a constant, which should be determined, and $\Lambda^{\mu \nu}$ is the energy-momentum tensor of a non-gravitational field obeying the energy-momentum conservation law 
\begin{equation}\label{eqA4}
{\Lambda^{\mu\nu}}_{;\nu} = 0 \,. 
\end{equation}

In the next, tensor $\Lambda^{\mu \nu}$ in Eq. (A3) will be interpreted as the result of the light-matter interaction described in Section Light-matter interaction. The constant $\psi$ standing at $\Lambda^{\mu \nu}$  in Eq. (A3) will be determined by applying the weak-field non-relativistic approximation, similarly as for determining the constant $8 \pi G/c^{4}$ standing at the energy-momentum tensor  $T^{\mu \nu}$.

Note that tensor $\Lambda^{\mu\nu}$ can formally be a part of the energy-momentum tensor $T^{\mu\nu}$. However, it is useful to treat it separately, in order to emphasize its non-gravitational nature similarly as done by Einstein in the case of the original cosmological constant $\Lambda$. In this way, tensor $T^{\mu\nu}$ is allocated for gravitational effects of mass and other physical fields only, but it does not reflect non-gravitational forces. Obviously, both approaches are mathematically equivalent, because if $\Lambda^{\mu\nu}$ is considered as a part of $T^{\mu\nu}$, matching the field equations for a weak non-relativistic field leads finally to decoupling of $\Lambda^{\mu\nu}$ and cancelling the factor $8 \pi G/c^4$  standing at the term with $\Lambda^{\mu\nu}$. 

For a perfect isotropic fluid, the energy-momentum tensor $T^{\mu\nu}$ reads
\begin{equation}\label{eqA5}
T^{\mu\nu} = \left(\rho + \frac{p}{c^2} \right) U^\mu U^\nu + pg^{\mu\nu}  \,, 
\end{equation}
where $\rho$ is the density, $p$ is the pressure, and $U^\mu$ is the four velocity. In analogy to (A5), the isotropic cosmological tensor $\Lambda^{\mu\nu}$ can be described as
\begin{equation}\label{eqA6}
\Lambda^{\mu\nu} = \left(\rho_\Lambda + \frac{p_\Lambda}{c^2} \right) U^\mu U^\nu + p_\Lambda g^{\mu\nu}  \,, 
\end{equation}
where $\rho_\Lambda$ is the density, $p_\Lambda$ is the pressure of the non-gravitational field  produced by the light-matter interaction. 

The unknown constant $\psi$ in Eq. (A3) can now be found in a straightforward way assuming the weak-field approximation and using Eq. (16). This equation can be split into the Poisson equations for the gravitational potential $\Phi_G$ and for the potential of the light-matter interaction $\Phi_\Lambda$ as follows
\begin{equation}\label{eqA7}
\Delta \Phi_G =  4 \pi G \rho \,,
\end{equation}
\begin{equation}\label{eqA8}
\Delta \Phi_\Lambda =  - \rho_\Lambda \,.
\end{equation}
Taking into account that $\Lambda^{00} = \rho_\Lambda c^2$ and applying exactly the same procedure as when determining the constant  $8 \pi G/c^{4}$  standing at tensor $T^{\mu \nu}$ in Eq. (A1), we get $\psi = 2/c^4$. Hence, Eq. (A3) finally reads
\begin{equation}\label{eqA9}
G^{\mu\nu} + \frac{2}{c^4} \Lambda^{\mu\nu} = \frac{8 \pi G}{c^4} T^{\mu\nu} \,. 
\end{equation}

Introducing the standard FLRW metric of the space defined by its Gaussian curvature $k$ and by the scale factor $a(t)$ \citep{Peacock1999, Ryden2016}
\begin{equation}\label{eqA10}
-c^2 d\tau^2 = -c^2 dt^2 + a^2(t) \left(\frac{dr^2}{1-kr^2} + r^2 d\Omega^2 \right)  \,, 
\end{equation}
in Eqs. (A9), (A5) and (A6), we get a modified form of the Friedmann equations, which involve effects of the non-gravitational field $\Lambda^{\mu\nu}$
\begin{equation}\label{eqA11}
{\left({\frac{\dot a}{a}}\right)}^2 = \frac{8\pi G}{3} \rho - \frac{2}{3} \rho_\Lambda - \frac{k c^2}{a^2} \,, 
\end{equation}
\begin{equation}\label{eqA12}
\frac{\ddot{a}}{a} = -\frac{4\pi G}{3} \left(\rho + \frac{3 p}{c^2} \right) + \frac{1}{3} \left(\rho_\Lambda + \frac{3 p_\Lambda}{c^2} \right)   \,. 
\end{equation}
Considering $\rho$ and $\rho_\Lambda$ depending on the scale factor $a(t)$ as $\rho = \rho_0 a^{-\alpha}$  and $\rho_\Lambda = \rho_{\Lambda 0} a^{-\beta}$, the equations of state for $T^{\mu\nu}$ and $\Lambda^{\mu\nu}$ yield
\begin{equation}\label{eqA13}
p = \frac{\alpha - 3}{3} c^2 \rho \,, 
\end{equation}
\begin{equation}\label{eqA14}
p_\Lambda = \frac{\beta - 3}{3} c^2 \rho_\Lambda \,, 
\end{equation}
and Eq. (A12) reads
\begin{equation}\label{eqA15}
\frac{\ddot{a}}{a} = -\frac{4\pi G}{3} \left(\alpha - 2 \right) \rho + \frac{1}{3} \left(\beta - 2\right) \rho_\Lambda   \,. 
\end{equation}
%


\bibliographystyle{aa}
\bibliography{paper}   

\begin{thebibliography}{163}
\expandafter\ifx\csname natexlab\endcsname\relax\def\natexlab#1{#1}\fi

\bibitem[{{Abbott} {et~al.}(2017){Abbott}, {Abbott}, {Abbott}, {Acernese},
  {Ackley}, {Adams}, {Adams}, {Addesso}, {Adhikari}, {Adya}, {Affeldt},
  {Afrough}, {Agarwal}, {Agathos}, {Agatsuma}, {Aggarwal}, {Aguiar}, {Aiello},
  {Ain}, {Ajith}, {Allen}, {Allen}, {Allocca}, {Altin}, {Amato}, {Ananyeva},
  {Anderson}, {Anderson}, {Angelova}, {Antier}, {Appert}, {Arai}, {Araya},
  {Areeda}, {Arnaud}, {Arun}, {Ascenzi}, {Ashton}, {Ast}, {Aston}, {Astone},
  {Atallah}, {Aufmuth}, {Aulbert}, {Aultoneal}, {Austin}, {Avila-Alvarez},
  {Babak}, {Bacon}, {Bader}, {Bae}, {Baker}, {Baldaccini}, {Ballardin},
  {Ballmer}, {Banagiri}, {Barayoga}, {Barclay}, {Barish}, {Barker}, {Barkett},
  {Barone}, {Barr}, {Barsotti}, {Barsuglia}, {Barta}, {Bartlett}, {Bartos},
  {Bassiri}, {Basti}, {Batch}, {Bawaj}, {Bayley}, {Bazzan}, {B{\'e}csy},
  {Beer}, {Bejger}, {Belahcene}, {Bell}, {Berger}, {Bergmann}, {Bero}, {Berry},
  {Bersanetti}, {Bertolini}, {Betzwieser}, {Bhagwat}, {Bhandare}, {Bilenko},
  {Billingsley}, {Billman}, {Birch}, {Birney}, {Birnholtz}, {Biscans},
  {Biscoveanu}, {Bisht}, {Bitossi}, {Biwer}, {Bizouard}, {Blackburn},
  {Blackman}, {Blair}, {Blair}, {Blair}, {Bloemen}, {Bock}, {Bode}, {Boer},
  {Bogaert}, {Bohe}, {Bondu}, {Bonilla}, {Bonnand}, {Boom}, {Bork}, {Boschi},
  {Bose}, {Bossie}, {Bouffanais}, {Bozzi}, {Bradaschia}, {Brady}, {Branchesi},
  {Brau}, {Briant}, {Brillet}, {Brinkmann}, {Brisson}, {Brockill}, {Broida},
  {Brooks}, {Brown}, {Brown}, {Brunett}, {Buchanan}, {Buikema}, {Bulik},
  {Bulten}, {Buonanno}, {Buskulic}, {Buy}, {Byer}, {Cabero}, {Cadonati},
  {Cagnoli}, {Cahillane}, {Bustillo}, {Callister}, {Calloni}, {Camp}, {Canepa},
  {Canizares}, {Cannon}, {Cao}, {Cao}, {Capano}, {Capocasa}, {Carbognani},
  {Caride}, {Carney}, {Diaz}, {Casentini}, {Caudill}, {Cavagli{\`a}},
  {Cavalier}, {Cavalieri}, {Cella}, {Cepeda}, {Cerd{\'a}-Dur{\'a}n},
  {Cerretani}, {Cesarini}, {Chamberlin}, {Chan}, {Chao}, {Charlton}, {Chase},
  {Chassand e-Mottin}, {Chatterjee}, {Chatziioannou}, {Cheeseboro}, {Chen},
  {Chen}, {Chen}, {Cheng}, {Chia}, {Chincarini}, {Chiummo}, {Chmiel}, {Cho},
  {Cho}, {Chow}, {Christensen}, {Chu}, {Chua}, {Chua}, {Chung}, {Chung},
  {Ciani}, {Ciolfi}, {Cirelli}, {Cirone}, {Clara}, {Clark}, {Clearwater},
  {Cleva}, {Cocchieri}, {Coccia}, {Cohadon}, {Cohen}, {Colla}, {Collette},
  {Cominsky}, {Constancio}, {Conti}, {Cooper}, {Corban}, {Corbitt},
  {Cordero-Carri{\'o}n}, {Corley}, {Cornish}, {Corsi}, {Cortese}, {Costa},
  {Coughlin}, {Coughlin}, {Coulon}, {Countryman}, {Couvares}, {Covas}, {Cowan},
  {Coward}, {Cowart}, {Coyne}, {Coyne}, {Creighton}, {Creighton}, {Cripe},
  {Crowder}, {Cullen}, {Cumming}, {Cunningham}, {Cuoco}, {Dal Canton},
  {D{\'a}lya}, {Danilishin}, {D'Antonio}, {Danzmann}, {Dasgupta}, {da Silva
  Costa}, {Datrier}, {Dattilo}, {Dave}, {Davier}, {Davis}, {Daw}, {Day}, {de},
  {Debra}, {Degallaix}, {de Laurentis}, {Del{\'e}glise}, {Del Pozzo}, {Demos},
  {Denker}, {Dent}, {de Pietri}, {Dergachev}, {De Rosa}, {Derosa}, {de Rossi},
  {Desalvo}, {de Varona}, {Devenson}, {Dhurandhar}, {D{\'\i}az}, {di Fiore},
  {di Giovanni}, {di Girolamo}, {di Lieto}, {di Pace}, {di Palma}, {di Renzo},
  {Doctor}, {Dolique}, {Donovan}, {Dooley}, {Doravari}, {Dorrington},
  {Douglas}, {Dovale {\'A}lvarez}, {Downes}, {Drago}, {Dreissigacker},
  {Driggers}, {Du}, {Ducrot}, {Dupej}, {Dwyer}, {Edo}, {Edwards}, {Effler},
  {Eggenstein}, {Ehrens}, {Eichholz}, {Eikenberry}, {Eisenstein}, {Essick},
  {Estevez}, {Etienne}, {Etzel}, {Evans}, {Evans}, {Factourovich}, {Fafone},
  {Fair}, {Fairhurst}, {Fan}, {Farinon}, {Farr}, {Farr}, {Fauchon-Jones},
  {Favata}, {Fays}, {Fee}, {Fehrmann}, {Feicht}, {Fejer}, {Fernandez-Galiana},
  {Ferrante}, {Ferreira}, {Ferrini}, {Fidecaro}, {Finstad}, {Fiori},
  {Fiorucci}, {Fishbach}, {Fisher}, {Fitz-Axen}, {Flaminio}, {Fletcher},
  {Fong}, {Font}, {Forsyth}, {Forsyth}, {Fournier}, {Frasca}, {Frasconi},
  {Frei}, {Freise}, {Frey}, {Frey}, {Fries}, {Fritschel}, {Frolov}, {Fulda},
  {Fyffe}, {Gabbard}, {Gadre}, {Gaebel}, {Gair}, {Gammaitoni}, {Ganija},
  {Gaonkar}, {Garcia-Quiros}, {Garufi}, {Gateley}, {Gaudio}, {Gaur},
  {Gayathri}, {Gehrels}, {Gemme}, {Genin}, {Gennai}, {George}, {George},
  {Gergely}, {Germain}, {Ghonge}, {Ghosh}, {Ghosh}, {Ghosh}, {Giaime},
  {Giardina}, {Giazotto}, {Gill}, {Glover}, {Goetz}, {Goetz}, {Gomes},
  {Goncharov}, {Gonz{\'a}lez}, {Castro}, {Gopakumar}, {Gorodetsky}, {Gossan},
  {Gosselin}, {Gouaty}, {Grado}, {Graef}, {Granata}, {Grant}, {Gras}, {Gray},
  {Greco}, {Green}, {Gretarsson}, {Groot}, {Grote}, {Grunewald}, {Gruning},
  {Guidi}, {Guo}, {Gupta}, {Gupta}, {Gushwa}, {Gustafson}, {Gustafson},
  {Halim}, {Hall}, {Hall}, {Hamilton}, {Hammond}, {Haney}, {Hanke}, {Hanks},
  {Hanna}, {Hannam}, {Hannuksela}, {Hanson}, {Hardwick}, {Harms}, {Harry},
  {Harry}, {Hart}, {Haster}, {Haughian}, {Healy}, {Heidmann}, {Heintze},
  {Heitmann}, {Hello}, {Hemming}, {Hendry}, {Heng}, {Hennig}, {Heptonstall},
  {Heurs}, {Hild}, {Hinderer}, {Hoak}, {Hofman}, {Holt}, {Holz}, {Hopkins},
  {Horst}, {Hough}, {Houston}, {Howell}, {Hreibi}, {Hu}, {Huerta}, {Huet},
  {Hughey}, {Husa}, {Huttner}, {Huynh-Dinh}, {Indik}, {Inta}, {Intini}, {Isa},
  {Isac}, {Isi}, {Iyer}, {Izumi}, {Jacqmin}, {Jani}, {Jaranowski}, {Jawahar},
  {Jim{\'e}nez-Forteza}, {Johnson}, {Jones}, {Jones}, {Jonker}, {Ju}, {Junker},
  {Kalaghatgi}, {Kalogera}, {Kamai}, {Kand hasamy}, {Kang}, {Kanner},
  {Kapadia}, {Karki}, {Karvinen}, {Kasprzack}, {Katolik}, {Katsavounidis},
  {Katzman}, {Kaufer}, {Kawabe}, {K{\'e}f{\'e}lian}, {Keitel}, {Kemball},
  {Kennedy}, {Kent}, {Key}, {Khalili}, {Khan}, {Khan}, {Khan}, {Khazanov},
  {Kijbunchoo}, {Kim}, {Kim}, {Kim}, {Kim}, {Kim}, {Kim}, {Kimbrell}, {King},
  {King}, {Kinley-Hanlon}, {Kirchhoff}, {Kissel}, {Kleybolte}, {Klimenko},
  {Knowles}, {Koch}, {Koehlenbeck}, {Koley}, {Kondrashov}, {Kontos}, {Korobko},
  {Korth}, {Kowalska}, {Kozak}, {Kr{\"a}mer}, {Kringel}, {Krishnan},
  {Kr{\'o}lak}, {Kuehn}, {Kumar}, {Kumar}, {Kumar}, {Kuo}, {Kutynia}, {Kwang},
  {Lackey}, {Lai}, {Landry}, {Lang}, {Lange}, {Lantz}, {Lanza},
  {Lartaux-Vollard}, {Lasky}, {Laxen}, {Lazzarini}, {Lazzaro}, {Leaci},
  {Leavey}, {Lee}, {Lee}, {Lee}, {Lee}, {Lee}, {Lehmann}, {Lenon}, {Leonardi},
  {Leroy}, {Letendre}, {Levin}, {Li}, {Linker}, {Littenberg}, {Liu}, {Liu},
  {Lo}, {Lockerbie}, {London}, {Lord}, {Lorenzini}, {Loriette}, {Lormand},
  {Losurdo}, {Lough}, {Lousto}, {Lovelace}, {L{\"u}ck}, {Lumaca}, {Lundgren},
  {Lynch}, {Ma}, {Macas}, {Macfoy}, {Machenschalk}, {Macinnis}, {MacLeod},
  {Hernandez}, {Maga{\~n}a-Sand oval}, {Zertuche}, {Magee}, {Majorana},
  {Maksimovic}, {Man}, {Mandic}, {Mangano}, {Mansell}, {Manske}, {Mantovani},
  {Marchesoni}, {Marion}, {M{\'a}rka}, {M{\'a}rka}, {Markakis}, {Markosyan},
  {Markowitz}, {Maros}, {Marquina}, {Martelli}, {Martellini}, {Martin},
  {Martin}, {Martynov}, {Mason}, {Massera}, {Masserot}, {Massinger},
  {Masso-Reid}, {Mastrogiovanni}, {Matas}, {Matichard}, {Matone}, {Mavalvala},
  {Mazumder}, {McCarthy}, {McClelland}, {McCormick}, {McCuller}, {McGuire},
  {McIntyre}, {McIver}, {McManus}, {McNeill}, {McRae}, {McWilliams}, {Meacher},
  {Meadors}, {Mehmet}, {Meidam}, {Mejuto-Villa}, {Melatos}, {Mendell},
  {Mercer}, {Merilh}, {Merzougui}, {Meshkov}, {Messenger}, {Messick},
  {Metzdorff}, {Meyers}, {Miao}, {Michel}, {Middleton}, {Mikhailov}, {Milano},
  {Miller}, {Miller}, {Miller}, {Millhouse}, {Milovich-Goff}, {Minazzoli},
  {Minenkov}, {Ming}, {Mishra}, {Mitra}, {Mitrofanov}, {Mitselmakher},
  {Mittleman}, {Moffa}, {Moggi}, {Mogushi}, {Mohan}, {Mohapatra}, {Montani},
  {Moore}, {Moraru}, {Moreno}, {Morriss}, {Mours}, {Mow-Lowry}, {Mueller},
  {Muir}, {Mukherjee}, {Mukherjee}, {Mukherjee}, {Mukund}, {Mullavey}, {Munch},
  {Mu{\~n}iz}, {Muratore}, {Murray}, {Napier}, {Nardecchia}, {Naticchioni},
  {Nayak}, {Neilson}, {Nelemans}, {Nelson}, {Nery}, {Neunzert}, {Nevin},
  {Newport}, {Newton}, {Ng}, {Nguyen}, {Nichols}, {Nielsen}, {Nissanke},
  {Nitz}, {Noack}, {Nocera}, {Nolting}, {North}, {Nuttall}, {Oberling},
  {O'Dea}, {Ogin}, {Oh}, {Oh}, {Ohme}, {Okada}, {Oliver}, {Oppermann}, {Oram},
  {O'Reilly}, {Ormiston}, {Ortega}, {O'Shaughnessy}, {Ossokine}, {Ottaway},
  {Overmier}, {Owen}, {Pace}, {Page}, {Page}, {Pai}, {Pai}, {Palamos},
  {Palashov}, {Palomba}, {Pal-Singh}, {Pan}, {Pan}, {Pang}, {Pang}, {Pankow},
  {Pannarale}, {Pant}, {Paoletti}, {Paoli}, {Papa}, {Parida}, {Parker},
  {Pascucci}, {Pasqualetti}, {Passaquieti}, {Passuello}, {Patil}, {Patricelli},
  {Pearlstone}, {Pedraza}, {Pedurand}, {Pekowsky}, {Pele}, {Penn}, {Perez},
  {Perreca}, {Perri}, {Pfeiffer}, {Phelps}, {Piccinni}, {Pichot},
  {Piergiovanni}, {Pierro}, {Pillant}, {Pinard}, {Pinto}, {Pirello}, {Pitkin},
  {Poe}, {Poggiani}, {Popolizio}, {Porter}, {Post}, {Powell}, {Prasad},
  {Pratt}, {Pratten}, {Predoi}, {Prestegard}, {Prijatelj}, {Principe},
  {Privitera}, {Prodi}, {Prokhorov}, {Puncken}, {Punturo}, {Puppo},
  {P{\"u}rrer}, {Qi}, {Quetschke}, {Quintero}, {Quitzow-James}, {Raab},
  {Rabeling}, {Radkins}, {Raffai}, {Raja}, {Rajan}, {Rajbhandari}, {Rakhmanov},
  {Ramirez}, {Ramos-Buades}, {Rapagnani}, {Raymond}, {Razzano}, {Read},
  {Regimbau}, {Rei}, {Reid}, {Reitze}, {Ren}, {Reyes}, {Ricci}, {Ricker},
  {Rieger}, {Riles}, {Rizzo}, {Robertson}, {Robie}, {Robinet}, {Rocchi},
  {Rolland}, {Rollins}, {Roma}, {Romano}, {Romano}, {Romel}, {Romie},
  {Rosi{\'n}ska}, {Ross}, {Rowan}, {R{\"u}diger}, {Ruggi}, {Rutins}, {Ryan},
  {Sachdev}, {Sadecki}, {Sadeghian}, {Sakellariadou}, {Salconi}, {Saleem},
  {Salemi}, {Samajdar}, {Sammut}, {Sampson}, {Sanchez}, {Sanchez},
  {Sanchis-Gual}, {Sand berg}, {Sanders}, {Sassolas}, {Sathyaprakash},
  {Saulson}, {Sauter}, {Savage}, {Sawadsky}, {Schale}, {Scheel}, {Scheuer},
  {Schmidt}, {Schmidt}, {Schnabel}, {Schofield}, {Sch{\"o}nbeck}, {Schreiber},
  {Schuette}, {Schulte}, {Schutz}, {Schwalbe}, {Scott}, {Scott}, {Seidel},
  {Sellers}, {Sengupta}, {Sentenac}, {Sequino}, {Sergeev}, {Shaddock},
  {Shaffer}, {Shah}, {Shahriar}, {Shaner}, {Shao}, {Shapiro}, {Shawhan},
  {Sheperd}, {Shoemaker}, {Shoemaker}, {Siellez}, {Siemens}, {Sieniawska},
  {Sigg}, {Silva}, {Singer}, {Singh}, {Singhal}, {Sintes}, {Slagmolen},
  {Smith}, {Smith}, {Smith}, {Somala}, {Son}, {Sonnenberg}, {Sorazu},
  {Sorrentino}, {Souradeep}, {Spencer}, {Srivastava}, {Staats}, {Staley},
  {Steer}, {Steinke}, {Steinlechner}, {Steinlechner}, {Steinmeyer},
  {Stevenson}, {Stone}, {Stops}, {Strain}, {Stratta}, {Strigin}, {Strunk},
  {Sturani}, {Stuver}, {Summerscales}, {Sun}, {Sunil}, {Suresh}, {Sutton},
  {Swinkels}, {Szczepa{\'n}czyk}, {Tacca}, {Tait}, {Talbot}, {Talukder},
  {Tanner}, {T{\'a}pai}, {Taracchini}, {Tasson}, {Taylor}, {Taylor}, {Tewari},
  {Theeg}, {Thies}, {Thomas}, {Thomas}, {Thomas}, {Thorne}, {Thrane}, {Tiwari},
  {Tiwari}, {Tokmakov}, {Toland}, {Tonelli}, {Tornasi}, {Torres-Forn{\'e}},
  {Torrie}, {T{\"o}yr{\"a}}, {Travasso}, {Traylor}, {Trinastic}, {Tringali},
  {Trozzo}, {Tsang}, {Tse}, {Tso}, {Tsukada}, {Tsuna}, {Tuyenbayev}, {Ueno},
  {Ugolini}, {Unnikrishnan}, {Urban}, {Usman}, {Vahlbruch}, {Vajente},
  {Valdes}, {van Bakel}, {van Beuzekom}, {van den Brand}, {van den Broeck},
  {Vander-Hyde}, {van der Schaaf}, {van Heijningen}, {van Veggel}, {Vardaro},
  {Varma}, {Vass}, {Vas{\'u}th}, {Vecchio}, {Vedovato}, {Veitch}, {Veitch},
  {Venkateswara}, {Venugopalan}, {Verkindt}, {Vetrano}, {Vicer{\'e}}, {Viets},
  {Vinciguerra}, {Vine}, {Vinet}, {Vitale}, {Vo}, {Vocca}, {Vorvick},
  {Vyatchanin}, {Wade}, {Wade}, {Wade}, {Walet}, {Walker}, {Wallace}, {Walsh},
  {Wang}, {Wang}, {Wang}, {Wang}, {Wang}, {Ward}, {Warner}, {Was}, {Watchi},
  {Weaver}, {Wei}, {Weinert}, {Weinstein}, {Weiss}, {Wen}, {Wessel},
  {We{\ss}els}, {Westerweck}, {Westphal}, {Wette}, {Whelan}, {Whitcomb},
  {Whiting}, {Whittle}, {Wilken}, {Williams}, {Williams}, {Williamson},
  {Willis}, {Willke}, {Wimmer}, {Winkler}, {Wipf}, {Wittel}, {Woan}, {Woehler},
  {Wofford}, {Wong}, {Worden}, {Wright}, {Wu}, {Wysocki}, {Xiao}, {Yamamoto},
  {Yancey}, {Yang}, {Yap}, {Yazback}, {Yu}, {Yu}, {Yvert}, {Zadro{\.z}ny},
  {Zanolin}, {Zelenova}, {Zendri}, {Zevin}, {Zhang}, {Zhang}, {Zhang}, {Zhang},
  {Zhao}, {Zhou}, {Zhou}, {Zhu}, {Zhu}, {Zimmerman}, {Zucker}, {Zweizig},
  {Foley}, {Coulter}, {Drout}, {Kasen}, {Kilpatrick}, {Madore},
  {Murguia-Berthier}, {Pan}, {Piro}, {Prochaska}, {Ramirez-Ruiz}, {Rest},
  {Rojas-Bravo}, {Shappee}, {Siebert}, {Simon}, {Ulloa}, {Annis},
  {Soares-Santos}, {Brout}, {Scolnic}, {Diehl}, {Frieman}, {Berger},
  {Alexander}, {Allam}, {Balbinot}, {Blanchard}, {Butler}, {Chornock}, {Cook},
  {Cowperthwaite}, {Drlica-Wagner}, {Drout}, {Durret}, {Eftekhari}, {Finley},
  {Fong}, {Fryer}, {Garc{\'\i}a-Bellido}, {Gill}, {Gruendl}, {Hanna},
  {Hartley}, {Herner}, {Huterer}, {Kasen}, {Kessler}, {Li}, {Lin}, {Lopes},
  {Louren{\c{c}}o}, {Margutti}, {Marriner}, {Marshall}, {Matheson}, {Medina},
  {Metzger}, {Mu{\~n}oz}, {Muir}, {Nicholl}, {Nugent}, {Palmese},
  {Paz-Chinch{\'o}n}, {Quataert}, {Sako}, {Sauseda}, {Schlegel}, {Secco},
  {Smith}, {Sobreira}, {Stebbins}, {Villar}, {Vivas}, {Wester}, {Williams},
  {Yanny}, {Zenteno}, {Abbott}, {Abdalla}, {Bechtol}, {Benoit-L{\'e}vy},
  {Bertin}, {Bridle}, {Brooks}, {Buckley-Geer}, {Burke}, {Rosell}, {Kind},
  {Carretero}, {Castander}, {Cunha}, {D'Andrea}, {da Costa}, {Davis}, {Depoy},
  {Desai}, {Dietrich}, {Estrada}, {Fernandez}, {Flaugher}, {Fosalba},
  {Gaztanaga}, {Gerdes}, {Giannantonio}, {Goldstein}, {Gruen}, {Gutierrez},
  {Hartley}, {Honscheid}, {Jain}, {James}, {Jeltema}, {Johnson}, {Kent},
  {Krause}, {Kron}, {Kuehn}, {Kuhlmann}, {Kuropatkin}, {Lahav}, {Lima}, {Maia},
  {March}, {Miller}, {Miquel}, {Neilsen}, {Nord}, {Ogando}, {Plazas}, {Romer},
  {Roodman}, {Rykoff}, {Sanchez}, {Scarpine}, {Schubnell}, {Sevilla-Noarbe},
  {Smith}, {Smith}, {Suchyta}, {Tarle}, {Thomas}, {Thomas}, {Troxel}, {Tucker},
  {Vikram}, {Walker}, {Weller}, {Zhang}, {Haislip}, {Kouprianov}, {Reichart},
  {Tartaglia}, {Sand}, {Valenti}, {Yang}, {Arcavi}, {Hosseinzadeh}, {Howell},
  {McCully}, {Poznanski}, {Vasylyev}, {Tanvir}, {Levan}, {Hjorth}, {Cano},
  {Copperwheat}, {de Ugarte-Postigo}, {Evans}, {Fynbo},
  {Gonz{\'a}lez-Fern{\'a}ndez}, {Greiner}, {Irwin}, {Lyman}, {Mandel},
  {McMahon}, {Milvang-Jensen}, {O'Brien}, {Osborne}, {Perley}, {Pian},
  {Palazzi}, {Rol}, {Rosetti}, {Rosswog}, {Rowlinson}, {Schulze}, {Steeghs},
  {Th{\"o}ne}, {Ulaczyk}, {Watson}, {Wiersema}, {Lipunov}, {Gorbovskoy},
  {Kornilov}, {Tyurina}, {Balanutsa}, {Vlasenko}, {Gorbunov}, {Podesta},
  {Levato}, {Saffe}, {Buckley}, {Budnev}, {Gress}, {Yurkov}, {Rebolo}, \&
  {Serra-Ricart}}]{Abbott2017}
{Abbott}, B.~P., {Abbott}, R., {Abbott}, T.~D., {et~al.} 2017, \nat, 551, 85

\bibitem[{{Aguirre}(1999{\natexlab{a}})}]{Aguirre1999a}
{Aguirre}, A. 1999{\natexlab{a}}, \apj, 525, 583

\bibitem[{{Aguirre} \& {Haiman}(2000)}]{Aguirre_Haiman2000}
{Aguirre}, A. \& {Haiman}, Z. 2000, \apj, 532, 28

\bibitem[{{Aguirre}(1999{\natexlab{b}})}]{Aguirre1999b}
{Aguirre}, A.~N. 1999{\natexlab{b}}, \apj, 512, L19

\bibitem[{{Aguirre}(2000)}]{Aguirre2000}
{Aguirre}, A.~N. 2000, \apj, 533, 1

\bibitem[{{Angl{\'e}s-Alc{\'a}zar} {et~al.}(2017){Angl{\'e}s-Alc{\'a}zar},
  {Faucher-Gigu{\`e}re}, {Kere{\v s}}, {Hopkins}, {Quataert}, \&
  {Murray}}]{Angles-Alcazar2017}
{Angl{\'e}s-Alc{\'a}zar}, D., {Faucher-Gigu{\`e}re}, C.-A., {Kere{\v s}}, D.,
  {et~al.} 2017, \mnras, 470, 4698

\bibitem[{{Bahcall} {et~al.}(1995){Bahcall}, {Lubin}, \&
  {Dorman}}]{Bahcall1995}
{Bahcall}, N.~A., {Lubin}, L.~M., \& {Dorman}, V. 1995, \apj, 447, L81

\bibitem[{{Bernstein}(2007)}]{Bernstein2007}
{Bernstein}, R.~A. 2007, \apj, 666, 663

\bibitem[{{Bernstein} {et~al.}(2002{\natexlab{a}}){Bernstein}, {Freedman}, \&
  {Madore}}]{Bernstein2002a}
{Bernstein}, R.~A., {Freedman}, W.~L., \& {Madore}, B.~F. 2002{\natexlab{a}},
  \apj, 571, 56

\bibitem[{{Bernstein} {et~al.}(2002{\natexlab{b}}){Bernstein}, {Freedman}, \&
  {Madore}}]{Bernstein2002b}
{Bernstein}, R.~A., {Freedman}, W.~L., \& {Madore}, B.~F. 2002{\natexlab{b}},
  \apj, 571, 85

\bibitem[{{Bernstein} {et~al.}(2002{\natexlab{c}}){Bernstein}, {Freedman}, \&
  {Madore}}]{Bernstein2002c}
{Bernstein}, R.~A., {Freedman}, W.~L., \& {Madore}, B.~F. 2002{\natexlab{c}},
  \apj, 571, 107

\bibitem[{{Bianchi} \& {Ferrara}(2005)}]{Bianchi_Ferrara2005}
{Bianchi}, S. \& {Ferrara}, A. 2005, \mnras, 358, 379

\bibitem[{{Bielby} {et~al.}(2017){Bielby}, {Crighton}, {Fumagalli}, {Morris},
  {Stott}, {Tejos}, \& {Cantalupo}}]{Bielby2017}
{Bielby}, R., {Crighton}, N.~H.~M., {Fumagalli}, M., {et~al.} 2017, \mnras,
  468, 1373

\bibitem[{{Birkinshaw}(1999)}]{Birkinshaw1999}
{Birkinshaw}, M. 1999, \physrep, 310, 97

\bibitem[{{Blanton} {et~al.}(2001){Blanton}, {Dalcanton}, {Eisenstein},
  {Loveday}, {Strauss}, {SubbaRao}, {Weinberg}, {Anderson}, {Annis}, {Bahcall},
  {Bernardi}, {Brinkmann}, {Brunner}, {Burles}, {Carey}, {Castander},
  {Connolly}, {Csabai}, {Doi}, {Finkbeiner}, {Friedman}, {Frieman}, {Fukugita},
  {Gunn}, {Hennessy}, {Hindsley}, {Hogg}, {Ichikawa}, {Ivezi{\'c}}, {Kent},
  {Knapp}, {Lamb}, {Leger}, {Long}, {Lupton}, {McKay}, {Meiksin}, {Merelli},
  {Munn}, {Narayanan}, {Newcomb}, {Nichol}, {Okamura}, {Owen}, {Pier}, {Pope},
  {Postman}, {Quinn}, {Rockosi}, {Schlegel}, {Schneider}, {Shimasaku},
  {Siegmund}, {Smee}, {Snir}, {Stoughton}, {Stubbs}, {Szalay}, {Szokoly},
  {Thakar}, {Tremonti}, {Tucker}, {Uomoto}, {Vanden Berk}, {Vogeley},
  {Waddell}, {Yanny}, {Yasuda}, \& {York}}]{Blanton2001}
{Blanton}, M.~R., {Dalcanton}, J., {Eisenstein}, D., {et~al.} 2001, \aj, 121,
  2358

\bibitem[{{Blanton} {et~al.}(2003){Blanton}, {Hogg}, {Bahcall}, {Brinkmann},
  {Britton}, {Connolly}, {Csabai}, {Fukugita}, {Loveday}, {Meiksin}, {Munn},
  {Nichol}, {Okamura}, {Quinn}, {Schneider}, {Shimasaku}, {Strauss}, {Tegmark},
  {Vogeley}, \& {Weinberg}}]{Blanton2003}
{Blanton}, M.~R., {Hogg}, D.~W., {Bahcall}, N.~A., {et~al.} 2003, \apj, 592,
  819

\bibitem[{{Bohlin} {et~al.}(1978){Bohlin}, {Savage}, \& {Drake}}]{Bohlin1978}
{Bohlin}, R.~C., {Savage}, B.~D., \& {Drake}, J.~F. 1978, \apj, 224, 132

\bibitem[{{Bonamente} {et~al.}(2006){Bonamente}, {Joy}, {LaRoque}, {Carlstrom},
  {Reese}, \& {Dawson}}]{Bonamente2006}
{Bonamente}, M., {Joy}, M.~K., {LaRoque}, S.~J., {et~al.} 2006, \apj, 647, 25

\bibitem[{{Bond} {et~al.}(2013){Bond}, {Nelan}, {VandenBerg}, {Schaefer}, \&
  {Harmer}}]{Bond2013}
{Bond}, H.~E., {Nelan}, E.~P., {VandenBerg}, D.~A., {Schaefer}, G.~H., \&
  {Harmer}, D. 2013, \apjl, 765, L12

\bibitem[{{Bond} {et~al.}(1991){Bond}, {Carr}, \& {Hogan}}]{Bond1991}
{Bond}, J.~R., {Carr}, B.~J., \& {Hogan}, C.~J. 1991, \apj, 367, 420

\bibitem[{{Bonvin} {et~al.}(2017){Bonvin}, {Courbin}, {Suyu}, {Marshall},
  {Rusu}, {Sluse}, {Tewes}, {Wong}, {Collett}, {Fassnacht}, {Treu}, {Auger},
  {Hilbert}, {Koopmans}, {Meylan}, {Rumbaugh}, {Sonnenfeld}, \&
  {Spiniello}}]{Bonvin2017}
{Bonvin}, V., {Courbin}, F., {Suyu}, S.~H., {et~al.} 2017, \mnras, 465, 4914

\bibitem[{{Bouwens} {et~al.}(2014{\natexlab{a}}){Bouwens}, {Bradley}, {Zitrin},
  {Coe}, {Franx}, {Zheng}, {Smit}, {Host}, {Postman}, {Moustakas}, {Labb{\'e}},
  {Carrasco}, {Molino}, {Donahue}, {Kelson}, {Meneghetti}, {Ben{\'{\i}}tez},
  {Lemze}, {Umetsu}, {Broadhurst}, {Moustakas}, {Rosati}, {Jouvel},
  {Bartelmann}, {Ford}, {Graves}, {Grillo}, {Infante}, {Jimenez-Teja}, {Lahav},
  {Maoz}, {Medezinski}, {Melchior}, {Merten}, {Nonino}, {Ogaz}, \&
  {Seitz}}]{Bouwens2014b}
{Bouwens}, R.~J., {Bradley}, L., {Zitrin}, A., {et~al.} 2014{\natexlab{a}},
  \apj, 795, 126

\bibitem[{{Bouwens} {et~al.}(2014{\natexlab{b}}){Bouwens}, {Illingworth},
  {Oesch}, {Labb{\'e}}, {van Dokkum}, {Trenti}, {Franx}, {Smit}, {Gonzalez}, \&
  {Magee}}]{Bouwens2014a}
{Bouwens}, R.~J., {Illingworth}, G.~D., {Oesch}, P.~A., {et~al.}
  2014{\natexlab{b}}, \apj, 793, 115

\bibitem[{{Brennan} {et~al.}(2018){Brennan}, {Choi}, {Somerville},
  {Hirschmann}, {Naab}, \& {Ostriker}}]{Brennan2018}
{Brennan}, R., {Choi}, E., {Somerville}, R.~S., {et~al.} 2018, \apj, 860, 14

\bibitem[{{Brown} {et~al.}(2001){Brown}, {Geller}, {Fabricant}, \&
  {Kurtz}}]{Brown2001}
{Brown}, W.~R., {Geller}, M.~J., {Fabricant}, D.~G., \& {Kurtz}, M.~J. 2001,
  \aj, 122, 714

\bibitem[{{Buchert} {et~al.}(2016){Buchert}, {Coley}, {Kleinert}, {Roukema}, \&
  {Wiltshire}}]{Buchert2016}
{Buchert}, T., {Coley}, A.~A., {Kleinert}, H., {Roukema}, B.~F., \&
  {Wiltshire}, D.~L. 2016, International Journal of Modern Physics D, 25,
  1630007

\bibitem[{{Bullock} \& {Boylan-Kolchin}(2017)}]{Bullock_Boylan-Kolchin2017}
{Bullock}, J.~S. \& {Boylan-Kolchin}, M. 2017, Annual Review of Astronomy and
  Astrophysics, 55, 343

\bibitem[{{Calzetti}(2001)}]{Calzetti2001}
{Calzetti}, D. 2001, \pasp, 113, 1449

\bibitem[{{Calzetti} {et~al.}(2000){Calzetti}, {Armus}, {Bohlin}, {Kinney},
  {Koornneef}, \& {Storchi-Bergmann}}]{Calzetti2000}
{Calzetti}, D., {Armus}, L., {Bohlin}, R.~C., {et~al.} 2000, \apj, 533, 682

\bibitem[{{Cardelli} {et~al.}(1989){Cardelli}, {Clayton}, \&
  {Mathis}}]{Cardelli1989}
{Cardelli}, J.~A., {Clayton}, G.~C., \& {Mathis}, J.~S. 1989, \apj, 345, 245

\bibitem[{{Carlstrom} {et~al.}(2002){Carlstrom}, {Holder}, \&
  {Reese}}]{Carlstrom2002}
{Carlstrom}, J.~E., {Holder}, G.~P., \& {Reese}, E.~D. 2002, Annual Review of
  Astronomy and Astrophysics, 40, 643

\bibitem[{{Chelouche} {et~al.}(2007){Chelouche}, {Koester}, \&
  {Bowen}}]{Chelouche2007}
{Chelouche}, D., {Koester}, B.~P., \& {Bowen}, D.~V. 2007, \apjl, 671, L97

\bibitem[{{Clarkson} {et~al.}(2007){Clarkson}, {Cort{\^e}s}, \&
  {Bassett}}]{Clarkson2007}
{Clarkson}, C., {Cort{\^e}s}, M., \& {Bassett}, B. 2007, \jcap, 2007, 011

\bibitem[{{Cooray}(2016)}]{Cooray2016}
{Cooray}, A. 2016, Royal Society Open Science, 3, 150555

\bibitem[{{Corasaniti}(2006)}]{Corasaniti2006}
{Corasaniti}, P.~S. 2006, \mnras, 372, 191

\bibitem[{{Cross} {et~al.}(2001){Cross}, {Driver}, {Couch}, {Baugh},
  {Bland-Hawthorn}, {Bridges}, {Cannon}, {Cole}, {Colless}, {Collins},
  {Dalton}, {Deeley}, {De Propris}, {Efstathiou}, {Ellis}, {Frenk},
  {Glazebrook}, {Jackson}, {Lahav}, {Lewis}, {Lumsden}, {Maddox}, {Madgwick},
  {Moody}, {Norberg}, {Peacock}, {Peterson}, {Price}, {Seaborne}, {Sutherland},
  {Tadros}, \& {Taylor}}]{Cross2001}
{Cross}, N., {Driver}, S.~P., {Couch}, W., {et~al.} 2001, \mnras, 324, 825

\bibitem[{{Cruz} {et~al.}(2005){Cruz}, {Mart{\'{\i}}nez-Gonz{\'a}lez},
  {Vielva}, \& {Cay{\'o}n}}]{Cruz2005}
{Cruz}, M., {Mart{\'{\i}}nez-Gonz{\'a}lez}, E., {Vielva}, P., \& {Cay{\'o}n},
  L. 2005, \mnras, 356, 29

\bibitem[{{Davies} {et~al.}(1998){Davies}, {Alton}, {Bianchi}, \&
  {Trewhella}}]{Davies1998}
{Davies}, J.~I., {Alton}, P., {Bianchi}, S., \& {Trewhella}, M. 1998, \mnras,
  300, 1006

\bibitem[{{Draine}(2003)}]{Draine2003}
{Draine}, B.~T. 2003, \araa, 41, 241

\bibitem[{{Draine}(2011)}]{Draine2011}
{Draine}, B.~T. 2011, {Physics of the Interstellar and Intergalactic Medium}

\bibitem[{{Draine} \& {Fraisse}(2009)}]{Draine_Fraisse2009}
{Draine}, B.~T. \& {Fraisse}, A.~A. 2009, \apj, 696, 1

\bibitem[{{Dwek} \& {Krennrich}(2005)}]{Dwek2005}
{Dwek}, E. \& {Krennrich}, F. 2005, \apj, 618, 657

\bibitem[{{Einstein}(1917)}]{Einstein1917}
{Einstein}, A. 1917, Sitzungsberichte der K{\"o}niglich Preu{\ss}ischen
  Akademie der Wissenschaften (Berlin, 142

\bibitem[{{Ellis} {et~al.}(2013){Ellis}, {McLure}, {Dunlop}, {Robertson},
  {Ono}, {Schenker}, {Koekemoer}, {Bowler}, {Ouchi}, {Rogers}, {Curtis-Lake},
  {Schneider}, {Charlot}, {Stark}, {Furlanetto}, \& {Cirasuolo}}]{Ellis2013}
{Ellis}, R.~S., {McLure}, R.~J., {Dunlop}, J.~S., {et~al.} 2013, \apjl, 763, L7

\bibitem[{{Ezquiaga} \&
  {Zumalac{\'a}rregui}(2017)}]{Ezquiaga_Zumalacarregui2017}
{Ezquiaga}, J.~M. \& {Zumalac{\'a}rregui}, M. 2017, \prl, 119, 251304

\bibitem[{{Fan} {et~al.}(2006){Fan}, {Strauss}, {Becker}, {White}, {Gunn},
  {Knapp}, {Richards}, {Schneider}, {Brinkmann}, \& {Fukugita}}]{Fan2006b}
{Fan}, X., {Strauss}, M.~A., {Becker}, R.~H., {et~al.} 2006, \aj, 132, 117

\bibitem[{{Finkelman} {et~al.}(2008){Finkelman}, {Brosch}, {Kniazev},
  {Buckley}, {O'Donoghue}, {Hashimoto}, {Loaring}, {Romero-Colmenero}, {Still},
  {Sefako}, \& {V{\"a}is{\"a}nen}}]{Finkelman2008}
{Finkelman}, I., {Brosch}, N., {Kniazev}, A.~Y., {et~al.} 2008, \mnras, 390,
  969

\bibitem[{{Flynn}(1994)}]{Flynn1994}
{Flynn}, G.~J. 1994, \planss, 42, 1151

\bibitem[{{Freedman} {et~al.}(2001){Freedman}, {Madore}, {Gibson}, {Ferrarese},
  {Kelson}, {Sakai}, {Mould}, {Kennicutt}, {Ford}, {Graham}, {Huchra},
  {Hughes}, {Illingworth}, {Macri}, \& {Stetson}}]{Freedman2001}
{Freedman}, W.~L., {Madore}, B.~F., {Gibson}, B.~K., {et~al.} 2001, \apj, 553,
  47

\bibitem[{{Freedman} {et~al.}(2019){Freedman}, {Madore}, {Hatt}, {Hoyt},
  {Jang}, {Beaton}, {Burns}, {Lee}, {Monson}, {Neeley}, {Phillips}, {Rich}, \&
  {Seibert}}]{Freedman2019}
{Freedman}, W.~L., {Madore}, B.~F., {Hatt}, D., {et~al.} 2019, \apj, 882, 34

\bibitem[{{Fujimoto} {et~al.}(2019){Fujimoto}, {Ouchi}, {Ferrara},
  {Pallottini}, {Ivison}, {Behrens}, {Gallerani}, {Arata}, {Yajima}, \&
  {Nagamine}}]{Fujimoto2019}
{Fujimoto}, S., {Ouchi}, M., {Ferrara}, A., {et~al.} 2019, \apj, 887, 107

\bibitem[{{Gilmore} {et~al.}(2012){Gilmore}, {Somerville}, {Primack}, \&
  {Dom{\'{\i}}nguez}}]{Gilmore2012}
{Gilmore}, R.~C., {Somerville}, R.~S., {Primack}, J.~R., \& {Dom{\'{\i}}nguez},
  A. 2012, \mnras, 422, 3189

\bibitem[{{Gold} {et~al.}(2011){Gold}, {Odegard}, {Weiland}, {Hill}, {Kogut},
  {Bennett}, {Hinshaw}, {Chen}, {Dunkley}, {Halpern}, {Jarosik}, {Komatsu},
  {Larson}, {Limon}, {Meyer}, {Nolta}, {Page}, {Smith}, {Spergel}, {Tucker},
  {Wollack}, \& {Wright}}]{Gold2011}
{Gold}, B., {Odegard}, N., {Weiland}, J.~L., {et~al.} 2011, \apjs, 192, 15

\bibitem[{{Gonz{\'a}lez} {et~al.}(2011){Gonz{\'a}lez}, {Labb{\'e}}, {Bouwens},
  {Illingworth}, {Franx}, \& {Kriek}}]{Gonzalez2011}
{Gonz{\'a}lez}, V., {Labb{\'e}}, I., {Bouwens}, R.~J., {et~al.} 2011, \apjl,
  735, L34

\bibitem[{{Hashimoto} {et~al.}(2018){Hashimoto}, {Laporte}, {Mawatari},
  {Ellis}, {Inoue}, {Zackrisson}, {Roberts-Borsani}, {Zheng}, {Tamura},
  {Bauer}, {Fletcher}, {Harikane}, {Hatsukade}, {Hayatsu}, {Matsuda}, {Matsuo},
  {Okamoto}, {Ouchi}, {Pell{\'o}}, {Rydberg}, {Shimizu}, {Taniguchi},
  {Umehata}, \& {Yoshida}}]{Hashimoto2018}
{Hashimoto}, T., {Laporte}, N., {Mawatari}, K., {et~al.} 2018, \nat, 557, 392

\bibitem[{{Hatch}(2016)}]{Hatch2016}
{Hatch}, N. 2016, Science, 354, 1102

\bibitem[{{Hauser} \& {Dwek}(2001)}]{Hauser2001}
{Hauser}, M.~G. \& {Dwek}, E. 2001, \araa, 39, 249

\bibitem[{{Hirashita} \& {Inoue}(2019)}]{Hirashita_Inoue2019}
{Hirashita}, H. \& {Inoue}, A.~K. 2019, \mnras, 487, 961

\bibitem[{{Hoag} {et~al.}(2018){Hoag}, {Brada{\v{c}}}, {Brammer}, {Huang},
  {Treu}, {Mason}, {Castellano}, {Di Criscienzo}, {Jones}, {Kelly},
  {Pentericci}, {Ryan}, {Schmidt}, \& {Trenti}}]{Hoag2018}
{Hoag}, A., {Brada{\v{c}}}, M., {Brammer}, G., {et~al.} 2018, \apj, 854, 39

\bibitem[{{Holwerda} {et~al.}(2007){Holwerda}, {Draine}, {Gordon},
  {Gonz{\'a}lez}, {Calzetti}, {Thornley}, {Buckalew}, {Allen}, \& {van der
  Kruit}}]{Holwerda2007}
{Holwerda}, B.~W., {Draine}, B., {Gordon}, K.~D., {et~al.} 2007, \aj, 134, 2226

\bibitem[{{Holwerda} {et~al.}(2005){Holwerda}, {Gonzalez}, {Allen}, \& {van der
  Kruit}}]{Holwerda2005b}
{Holwerda}, B.~W., {Gonzalez}, R.~A., {Allen}, R.~J., \& {van der Kruit}, P.~C.
  2005, \aj, 129, 1381

\bibitem[{{Hopkins} {et~al.}(2012){Hopkins}, {Quataert}, \&
  {Murray}}]{Hopkins2012}
{Hopkins}, P.~F., {Quataert}, E., \& {Murray}, N. 2012, \mnras, 421, 3522

\bibitem[{{Howlett} \& {Davis}(2020)}]{Howlett_Davis2020}
{Howlett}, C. \& {Davis}, T.~M. 2020, \mnras, 492, 3803

\bibitem[{{Ichiki}(2014)}]{Ichiki2014}
{Ichiki}, K. 2014, Progress of Theoretical and Experimental Physics, 2014,
  06B109

\bibitem[{{Jackson}(2015)}]{Jackson2015}
{Jackson}, N. 2015, Living Reviews in Relativity, 18, 2

\bibitem[{{Jones} {et~al.}(1996){Jones}, {Tielens}, \&
  {Hollenbach}}]{Jones1996}
{Jones}, A.~P., {Tielens}, A.~G.~G.~M., \& {Hollenbach}, D.~J. 1996, \apj, 469,
  740

\bibitem[{{Kajisawa} {et~al.}(2009){Kajisawa}, {Ichikawa}, {Tanaka}, {Konishi},
  {Yamada}, {Akiyama}, {Suzuki}, {Tokoku}, {Uchimoto}, {Yoshikawa}, {Ouchi},
  {Iwata}, {Hamana}, \& {Onodera}}]{Kajisawa2009}
{Kajisawa}, M., {Ichikawa}, T., {Tanaka}, I., {et~al.} 2009, \apj, 702, 1393

\bibitem[{{Kippenhahn} {et~al.}(2012){Kippenhahn}, {Weigert}, \&
  {Weiss}}]{Kippenhahn2012}
{Kippenhahn}, R., {Weigert}, A., \& {Weiss}, A. 2012, {Stellar Structure and
  Evolution}

\bibitem[{{Kneiske} {et~al.}(2004){Kneiske}, {Bretz}, {Mannheim}, \&
  {Hartmann}}]{Kneiske2004}
{Kneiske}, T.~M., {Bretz}, T., {Mannheim}, K., \& {Hartmann}, D.~H. 2004, \aap,
  413, 807

\bibitem[{{Kocifaj} {et~al.}(1999){Kocifaj}, {Kapisinsky}, \&
  {Kundracik}}]{Kocifaj1999}
{Kocifaj}, M., {Kapisinsky}, I., \& {Kundracik}, F. 1999, \jqsrt, 63, 1

\bibitem[{{Kohout} {et~al.}(2014){Kohout}, {Kallonen}, {Suuronen}, {Rochette},
  {Hutzler}, {Gattacceca}, {Badjukov}, {Sk{\'a}La}, {B{\"o}Hmov{\'a}}, \&
  {{\v{C}}Uda}}]{Kohout2014}
{Kohout}, T., {Kallonen}, A., {Suuronen}, J.~P., {et~al.} 2014, Meteoritics and
  Planetary Science, 49, 1157

\bibitem[{{Kroupa}(2012)}]{Kroupa2012}
{Kroupa}, P. 2012, Publications of the Astronomical Society of Australia, 29,
  395

\bibitem[{{Kroupa}(2015)}]{Kroupa2015}
{Kroupa}, P. 2015, Canadian Journal of Physics, 93, 169

\bibitem[{{Lagache} {et~al.}(2005){Lagache}, {Puget}, \& {Dole}}]{Lagache2005}
{Lagache}, G., {Puget}, J.-L., \& {Dole}, H. 2005, \araa, 43, 727

\bibitem[{{Laporte} {et~al.}(2017){Laporte}, {Ellis}, {Boone}, {Bauer},
  {Qu{\'e}nard}, {Roberts-Borsani}, {Pell{\'o}}, {P{\'e}rez-Fournon}, \&
  {Streblyanska}}]{Laporte2017}
{Laporte}, N., {Ellis}, R.~S., {Boone}, F., {et~al.} 2017, \apjl, 837, L21

\bibitem[{{Lazarian} \& {Prunet}(2002)}]{Lazarian2002}
{Lazarian}, A. \& {Prunet}, S. 2002, in American Institute of Physics
  Conference Series, Vol. 609, Astrophysical Polarized Backgrounds, ed.
  S.~{Cecchini}, S.~{Cortiglioni}, R.~{Sault}, \& C.~{Sbarra}, 32--43

\bibitem[{{Lee} {et~al.}(2012){Lee}, {Ferguson}, {Wiklind}, {Dahlen},
  {Dickinson}, {Giavalisco}, {Grogin}, {Papovich}, {Messias}, {Guo}, \&
  {Lin}}]{Lee2012}
{Lee}, K.-S., {Ferguson}, H.~C., {Wiklind}, T., {et~al.} 2012, \apj, 752, 66

\bibitem[{{Li} {et~al.}(2016){Li}, {Wang}, {Liao}, \& {Zhu}}]{Li2016}
{Li}, Z., {Wang}, G.-J., {Liao}, K., \& {Zhu}, Z.-H. 2016, \apj, 833, 240

\bibitem[{{Liao}(2019)}]{Liao2019}
{Liao}, K. 2019, \apj, 885, 70

\bibitem[{{Lisenfeld} {et~al.}(2008){Lisenfeld}, {Rela{\~n}o},
  {V{\'{\i}}lchez}, {Battaner}, \& {Hermelo}}]{Lisenfeld2008}
{Lisenfeld}, U., {Rela{\~n}o}, M., {V{\'{\i}}lchez}, J., {Battaner}, E., \&
  {Hermelo}, I. 2008, in IAU Symposium, Vol. 255, IAU Symposium, ed. L.~K.
  {Hunt}, S.~C. {Madden}, \& R.~{Schneider}, 260--264

\bibitem[{{Madau} \& {Dickinson}(2014)}]{Madau_Dickinson2014}
{Madau}, P. \& {Dickinson}, M. 2014, \araa, 52, 415

\bibitem[{{Madau} \& {Pozzetti}(2000)}]{Madau2000}
{Madau}, P. \& {Pozzetti}, L. 2000, \mnras, 312, L9

\bibitem[{{Marchesini} {et~al.}(2009){Marchesini}, {van Dokkum}, {F{\"o}rster
  Schreiber}, {Franx}, {Labb{\'e}}, \& {Wuyts}}]{Marchesini2009}
{Marchesini}, D., {van Dokkum}, P.~G., {F{\"o}rster Schreiber}, N.~M., {et~al.}
  2009, \apj, 701, 1765

\bibitem[{{Martin}(2005)}]{Martin2005}
{Martin}, C.~L. 2005, \apj, 621, 227

\bibitem[{{Mathis}(1990)}]{Mathis1990}
{Mathis}, J.~S. 1990, \araa, 28, 37

\bibitem[{{Mathis} {et~al.}(1977){Mathis}, {Rumpl}, \&
  {Nordsieck}}]{Mathis1977}
{Mathis}, J.~S., {Rumpl}, W., \& {Nordsieck}, K.~H. 1977, \apj, 217, 425

\bibitem[{{McLure} {et~al.}(2013){McLure}, {Dunlop}, {Bowler}, {Curtis-Lake},
  {Schenker}, {Ellis}, {Robertson}, {Koekemoer}, {Rogers}, {Ono}, {Ouchi},
  {Charlot}, {Wild}, {Stark}, {Furlanetto}, {Cirasuolo}, \&
  {Targett}}]{McLure2013}
{McLure}, R.~J., {Dunlop}, J.~S., {Bowler}, R.~A.~A., {et~al.} 2013, \mnras,
  432, 2696

\bibitem[{{Meiksin}(2009)}]{Meiksin2009}
{Meiksin}, A.~A. 2009, Reviews of Modern Physics, 81, 1405

\bibitem[{{M{\'e}nard} \& {Fukugita}(2012)}]{Menard_Fukugita2012}
{M{\'e}nard}, B. \& {Fukugita}, M. 2012, \apj, 754, 116

\bibitem[{{M{\'e}nard} {et~al.}(2010{\natexlab{a}}){M{\'e}nard}, {Kilbinger},
  \& {Scranton}}]{Menard2010b}
{M{\'e}nard}, B., {Kilbinger}, M., \& {Scranton}, R. 2010{\natexlab{a}},
  \mnras, 406, 1815

\bibitem[{{M{\'e}nard} {et~al.}(2010{\natexlab{b}}){M{\'e}nard}, {Scranton},
  {Fukugita}, \& {Richards}}]{Menard2010a}
{M{\'e}nard}, B., {Scranton}, R., {Fukugita}, M., \& {Richards}, G.
  2010{\natexlab{b}}, \mnras, 405, 1025

\bibitem[{{Muller} {et~al.}(2008){Muller}, {Wu}, {Hsieh}, {Gonz{\'a}lez},
  {Loinard}, {Yee}, \& {Gladders}}]{Muller2008}
{Muller}, S., {Wu}, S.-Y., {Hsieh}, B.-C., {et~al.} 2008, \apj, 680, 975

\bibitem[{{Muratov} {et~al.}(2017){Muratov}, {Kere{\v{s}}},
  {Faucher-Gigu{\`e}re}, {Hopkins}, {Ma}, {Angl{\'e}s-Alc{\'a}zar}, {Chan},
  {Torrey}, {Hafen}, {Quataert}, \& {Murray}}]{Muratov2017}
{Muratov}, A.~L., {Kere{\v{s}}}, D., {Faucher-Gigu{\`e}re}, C.-A., {et~al.}
  2017, \mnras, 468, 4170

\bibitem[{{Muzahid} {et~al.}(2015){Muzahid}, {Kacprzak}, {Churchill},
  {Charlton}, {Nielsen}, {Mathes}, \& {Trujillo-Gomez}}]{Muzahid2015}
{Muzahid}, S., {Kacprzak}, G.~G., {Churchill}, C.~W., {et~al.} 2015, \apj, 811,
  132

\bibitem[{{Narlikar} {et~al.}(2003){Narlikar}, {Vishwakarma}, {Hajian},
  {Souradeep}, {Burbidge}, \& {Hoyle}}]{Narlikar2003}
{Narlikar}, J.~V., {Vishwakarma}, R.~G., {Hajian}, A., {et~al.} 2003, \apj,
  585, 1

\bibitem[{{Nicastro} {et~al.}(2018){Nicastro}, {Kaastra}, {Krongold},
  {Borgani}, {Branchini}, {Cen}, {Dadina}, {Danforth}, {Elvis}, {Fiore},
  {Gupta}, {Mathur}, {Mayya}, {Paerels}, {Piro}, {Rosa-Gonzalez}, {Schaye},
  {Shull}, {Torres-Zafra}, {Wijers}, \& {Zappacosta}}]{Nicastro2018}
{Nicastro}, F., {Kaastra}, J., {Krongold}, Y., {et~al.} 2018, \nat, 558, 406

\bibitem[{{Oesch} {et~al.}(2014){Oesch}, {Bouwens}, {Illingworth}, {Labb{\'e}},
  {Smit}, {Franx}, {van Dokkum}, {Momcheva}, {Ashby}, {Fazio}, {Huang},
  {Willner}, {Gonzalez}, {Magee}, {Trenti}, {Brammer}, {Skelton}, \&
  {Spitler}}]{Oesch2014}
{Oesch}, P.~A., {Bouwens}, R.~J., {Illingworth}, G.~D., {et~al.} 2014, \apj,
  786, 108

\bibitem[{{Oesch} {et~al.}(2018){Oesch}, {Bouwens}, {Illingworth}, {Labb{\'e}},
  \& {Stefanon}}]{Oesch2018}
{Oesch}, P.~A., {Bouwens}, R.~J., {Illingworth}, G.~D., {Labb{\'e}}, I., \&
  {Stefanon}, M. 2018, \apj, 855, 105

\bibitem[{{Oesch} {et~al.}(2016){Oesch}, {Brammer}, {van Dokkum},
  {Illingworth}, {Bouwens}, {Labb{\'e}}, {Franx}, {Momcheva}, {Ashby}, {Fazio},
  {Gonzalez}, {Holden}, {Magee}, {Skelton}, {Smit}, {Spitler}, {Trenti}, \&
  {Willner}}]{Oesch2016}
{Oesch}, P.~A., {Brammer}, G., {van Dokkum}, P.~G., {et~al.} 2016, \apj, 819,
  129

\bibitem[{{Peacock}(1999)}]{Peacock1999}
{Peacock}, J.~A. 1999, {Cosmological Physics}

\bibitem[{{Peek} {et~al.}(2015){Peek}, {M{\'e}nard}, \& {Corrales}}]{Peek2015}
{Peek}, J.~E.~G., {M{\'e}nard}, B., \& {Corrales}, L. 2015, \apj, 813, 7

\bibitem[{{Penton} {et~al.}(2002){Penton}, {Stocke}, \& {Shull}}]{Penton2002}
{Penton}, S.~V., {Stocke}, J.~T., \& {Shull}, J.~M. 2002, \apj, 565, 720

\bibitem[{{P{\'e}rez-Gonz{\'a}lez} {et~al.}(2008){P{\'e}rez-Gonz{\'a}lez},
  {Rieke}, {Villar}, {Barro}, {Blaylock}, {Egami}, {Gallego}, {Gil de Paz},
  {Pascual}, {Zamorano}, \& {Donley}}]{Perez_Gonzalez2008}
{P{\'e}rez-Gonz{\'a}lez}, P.~G., {Rieke}, G.~H., {Villar}, V., {et~al.} 2008,
  \apj, 675, 234

\bibitem[{{Perlmutter} {et~al.}(1999){Perlmutter}, {Aldering}, {Goldhaber},
  {Knop}, {Nugent}, {Castro}, {Deustua}, {Fabbro}, {Goobar}, {Groom}, {Hook},
  {Kim}, {Kim}, {Lee}, {Nunes}, {Pain}, {Pennypacker}, {Quimby}, {Lidman},
  {Ellis}, {Irwin}, {McMahon}, {Ruiz-Lapuente}, {Walton}, {Schaefer}, {Boyle},
  {Filippenko}, {Matheson}, {Fruchter}, {Panagia}, {Newberg}, {Couch}, \&
  {Project}}]{Perlmutter1999}
{Perlmutter}, S., {Aldering}, G., {Goldhaber}, G., {et~al.} 1999, \apj, 517,
  565

\bibitem[{{P{\'e}roux} {et~al.}(2003){P{\'e}roux}, {McMahon},
  {Storrie-Lombardi}, \& {Irwin}}]{Peroux2003}
{P{\'e}roux}, C., {McMahon}, R.~G., {Storrie-Lombardi}, L.~J., \& {Irwin},
  M.~J. 2003, \mnras, 346, 1103

\bibitem[{{Pessa} {et~al.}(2018){Pessa}, {Tejos}, {Barrientos}, {Werk},
  {Bielby}, {Padilla}, {Morris}, {Prochaska}, {Lopez}, \&
  {Hummels}}]{Pessa2018}
{Pessa}, I., {Tejos}, N., {Barrientos}, L.~F., {et~al.} 2018, \mnras, 477, 2991

\bibitem[{{Pettini}(2004)}]{Pettini2004}
{Pettini}, M. 2004, in Cosmochemistry. The melting pot of the elements, ed.
  C.~{Esteban}, R.~{Garc{\'{\i}}a L{\'o}pez}, A.~{Herrero}, \&
  F.~{S{\'a}nchez}, 257--298

\bibitem[{{Pettini} {et~al.}(2003){Pettini}, {Madau}, {Bolte}, {Prochaska},
  {Ellison}, \& {Fan}}]{Pettini2003}
{Pettini}, M., {Madau}, P., {Bolte}, M., {et~al.} 2003, \apj, 594, 695

\bibitem[{{Planck Collaboration} {et~al.}(2015){Planck Collaboration}, {Ade},
  {Aghanim}, {Alina}, {Alves}, {Armitage-Caplan}, {Arnaud}, {Arzoumanian},
  {Ashdown}, {Atrio-Barand ela}, {Aumont}, {Baccigalupi}, {Banday}, {Barreiro},
  {Battaner}, {Benabed}, {Benoit-L{\'e}vy}, {Bernard}, {Bersanelli},
  {Bielewicz}, {Bock}, {Bond}, {Borrill}, {Bouchet}, {Boulanger}, {Bracco},
  {Burigana}, {Butler}, {Cardoso}, {Catalano}, {Chamballu}, {Chary}, {Chiang},
  {Christensen}, {Colombi}, {Colombo}, {Combet}, {Couchot}, {Coulais}, {Crill},
  {Curto}, {Cuttaia}, {Danese}, {Davies}, {Davis}, {de Bernardis}, {de Gouveia
  Dal Pino}, {de Rosa}, {de Zotti}, {Delabrouille}, {D{\'e}sert}, {Dickinson},
  {Diego}, {Donzelli}, {Dor{\'e}}, {Douspis}, {Dunkley}, {Dupac}, {Efstathiou},
  {En{\ss}lin}, {Eriksen}, {Falgarone}, {Ferri{\`e}re}, {Finelli}, {Forni},
  {Frailis}, {Fraisse}, {Franceschi}, {Galeotta}, {Ganga}, {Ghosh}, {Giard},
  {Giraud-H{\'e}raud}, {Gonz{\'a}lez-Nuevo}, {G{\'o}rski}, {Gregorio},
  {Gruppuso}, {Guillet}, {Hansen}, {Harrison}, {Helou},
  {Hern{\'a}ndez-Monteagudo}, {Hildebrand t}, {Hivon}, {Hobson}, {Holmes},
  {Hornstrup}, {Huffenberger}, {Jaffe}, {Jaffe}, {Jones}, {Juvela},
  {Keih{\"a}nen}, {Keskitalo}, {Kisner}, {Kneissl}, {Knoche}, {Kunz},
  {Kurki-Suonio}, {Lagache}, {L{\"a}hteenm{\"a}ki}, {Lamarre}, {Lasenby},
  {Lawrence}, {Leahy}, {Leonardi}, {Levrier}, {Liguori}, {Lilje},
  {Linden-V{\o}rnle}, {L{\'o}pez-Caniego}, {Lubin}, {Mac{\'\i}as-P{\'e}rez},
  {Maffei}, {Magalh{\~a}es}, {Maino}, {Mandolesi}, {Maris}, {Marshall},
  {Martin}, {Mart{\'\i}nez-Gonz{\'a}lez}, {Masi}, {Matarrese}, {Mazzotta},
  {Melchiorri}, {Mendes}, {Mennella}, {Migliaccio}, {Miville-Desch{\^e}nes},
  {Moneti}, {Montier}, {Morgante}, {Mortlock}, {Munshi}, {Murphy}, {Naselsky},
  {Nati}, {Natoli}, {Netterfield}, {Noviello}, {Novikov}, {Novikov},
  {Oxborrow}, {Pagano}, {Pajot}, {Paladini}, {Paoletti}, {Pasian}, {Pearson},
  {Perdereau}, {Perotto}, {Perrotta}, {Piacentini}, {Piat}, {Pietrobon},
  {Plaszczynski}, {Poidevin}, {Pointecouteau}, {Polenta}, {Popa}, {Pratt},
  {Prunet}, {Puget}, {Rachen}, {Reach}, {Rebolo}, {Reinecke}, {Remazeilles},
  {Renault}, {Ricciardi}, {Riller}, {Ristorcelli}, {Rocha}, {Rosset},
  {Roudier}, {Rubi{\~n}o-Mart{\'\i}n}, {Rusholme}, {Sandri}, {Savini}, {Scott},
  {Spencer}, {Stolyarov}, {Stompor}, {Sudiwala}, {Sutton}, {Suur-Uski},
  {Sygnet}, {Tauber}, {Terenzi}, {Toffolatti}, {Tomasi}, {Tristram}, {Tucci},
  {Umana}, {Valenziano}, {Valiviita}, {Van Tent}, {Vielva}, {Villa}, {Wade},
  {Wandelt}, {Zacchei}, \& {Zonca}}]{Ade2015_Planck_XIX_Polarized_emission}
{Planck Collaboration}, {Ade}, P.~A.~R., {Aghanim}, N., {et~al.} 2015, \aap,
  576, A104

\bibitem[{{Planck Collaboration} {et~al.}(2014){Planck Collaboration}, {Ade},
  {Aghanim}, {Armitage-Caplan}, {Arnaud}, {Ashdown}, {Atrio-Barandela},
  {Aumont}, {Baccigalupi}, {Banday}, \&
  et~al.}]{Ade2014_Planck_XXIV_Primordial}
{Planck Collaboration}, {Ade}, P.~A.~R., {Aghanim}, N., {et~al.} 2014, \aap,
  571, A24

\bibitem[{{Planck Collaboration} {et~al.}(2016{\natexlab{a}}){Planck
  Collaboration}, {Ade}, {Aghanim}, {Arnaud}, {Ashdown}, {Aumont},
  {Baccigalupi}, {Banday}, {Barreiro}, {Bartlett}, {Bartolo}, {Battaner},
  {Battye}, {Benabed}, {Beno{\^\i}t}, {Benoit-L{\'e}vy}, {Bernard},
  {Bersanelli}, {Bielewicz}, {Bock}, {Bonaldi}, {Bonavera}, {Bond}, {Borrill},
  {Bouchet}, {Boulanger}, {Bucher}, {Burigana}, {Butler}, {Calabrese},
  {Cardoso}, {Catalano}, {Challinor}, {Chamballu}, {Chary}, {Chiang}, {Chluba},
  {Christensen}, {Church}, {Clements}, {Colombi}, {Colombo}, {Combet},
  {Coulais}, {Crill}, {Curto}, {Cuttaia}, {Danese}, {Davies}, {Davis}, {de
  Bernardis}, {de Rosa}, {de Zotti}, {Delabrouille}, {D{\'e}sert}, {Di
  Valentino}, {Dickinson}, {Diego}, {Dolag}, {Dole}, {Donzelli}, {Dor{\'e}},
  {Douspis}, {Ducout}, {Dunkley}, {Dupac}, {Efstathiou}, {Elsner},
  {En{\ss}lin}, {Eriksen}, {Farhang}, {Fergusson}, {Finelli}, {Forni},
  {Frailis}, {Fraisse}, {Franceschi}, {Frejsel}, {Galeotta}, {Galli}, {Ganga},
  {Gauthier}, {Gerbino}, {Ghosh}, {Giard}, {Giraud-H{\'e}raud}, {Giusarma},
  {Gjerl{\o}w}, {Gonz{\'a}lez-Nuevo}, {G{\'o}rski}, {Gratton}, {Gregorio},
  {Gruppuso}, {Gudmundsson}, {Hamann}, {Hansen}, {Hanson}, {Harrison}, {Helou},
  {Henrot- Versill{\'e}}, {Hern{\'a}ndez-Monteagudo}, {Herranz}, {Hildebrandt},
  {Hivon}, {Hobson}, {Holmes}, {Hornstrup}, {Hovest}, {Huang}, {Huffenberger},
  {Hurier}, {Jaffe}, {Jaffe}, {Jones}, {Juvela}, {Keih{\"a}nen}, {Keskitalo},
  {Kisner}, {Kneissl}, {Knoche}, {Knox}, {Kunz}, {Kurki-Suonio}, {Lagache},
  {L{\"a}hteenm{\"a}ki}, {Lamarre}, {Lasenby}, {Lattanzi}, {Lawrence}, {Leahy},
  {Leonardi}, {Lesgourgues}, {Levrier}, {Lewis}, {Liguori}, {Lilje},
  {Linden-V{\o}rnle}, {L{\'o}pez-Caniego}, {Lubin}, {Mac{\'\i}as-P{\'e}rez},
  {Maggio}, {Maino}, {Mandolesi}, {Mangilli}, {Marchini}, {Maris}, {Martin},
  {Martinelli}, {Mart{\'\i}nez-Gonz{\'a}lez}, {Masi}, {Matarrese}, {McGehee},
  {Meinhold}, {Melchiorri}, {Melin}, {Mendes}, {Mennella}, {Migliaccio},
  {Millea}, {Mitra}, {Miville-Desch{\^e}nes}, {Moneti}, {Montier}, {Morgante},
  {Mortlock}, {Moss}, {Munshi}, {Murphy}, {Naselsky}, {Nati}, {Natoli},
  {Netterfield}, {N{\o}rgaard-Nielsen}, {Noviello}, {Novikov}, {Novikov},
  {Oxborrow}, {Paci}, {Pagano}, {Pajot}, {Paladini}, {Paoletti}, {Partridge},
  {Pasian}, {Patanchon}, {Pearson}, {Perdereau}, {Perotto}, {Perrotta},
  {Pettorino}, {Piacentini}, {Piat}, {Pierpaoli}, {Pietrobon}, {Plaszczynski},
  {Pointecouteau}, {Polenta}, {Popa}, {Pratt}, {Pr{\'e}zeau}, {Prunet},
  {Puget}, {Rachen}, {Reach}, {Rebolo}, {Reinecke}, {Remazeilles}, {Renault},
  {Renzi}, {Ristorcelli}, {Rocha}, {Rosset}, {Rossetti}, {Roudier},
  {Rouill{\'e} d'Orfeuil}, {Rowan-Robinson}, {Rubi{\~n}o-Mart{\'\i}n},
  {Rusholme}, {Said}, {Salvatelli}, {Salvati}, {Sandri}, {Santos},
  {Savelainen}, {Savini}, {Scott}, {Seiffert}, {Serra}, {Shellard}, {Spencer},
  {Spinelli}, {Stolyarov}, {Stompor}, {Sudiwala}, {Sunyaev}, {Sutton},
  {Suur-Uski}, {Sygnet}, {Tauber}, {Terenzi}, {Toffolatti}, {Tomasi},
  {Tristram}, {Trombetti}, {Tucci}, {Tuovinen}, {T{\"u}rler}, {Umana},
  {Valenziano}, {Valiviita}, {Van Tent}, {Vielva}, {Villa}, {Wade}, {Wandelt},
  {Wehus}, {White}, {White}, {Wilkinson}, {Yvon}, {Zacchei}, \&
  {Zonca}}]{Ade2015_Planck_XIII_Cosmological_parameters}
{Planck Collaboration}, {Ade}, P.~A.~R., {Aghanim}, N., {et~al.}
  2016{\natexlab{a}}, \aap, 594, A13

\bibitem[{{Planck Collaboration} {et~al.}(2018){Planck Collaboration},
  {Aghanim}, {Akrami}, {Ashdown}, {Aumont}, {Baccigalupi}, {Ballardini},
  {Banday}, {Barreiro}, {Bartolo}, {Basak}, {Battye}, {Benabed}, {Bernard},
  {Bersanelli}, {Bielewicz}, {Bock}, {Bond}, {Borrill}, {Bouchet}, {Boulanger},
  {Bucher}, {Burigana}, {Butler}, {Calabrese}, {Cardoso}, {Carron},
  {Challinor}, {Chiang}, {Chluba}, {Colombo}, {Combet}, {Contreras}, {Crill},
  {Cuttaia}, {de Bernardis}, {de Zotti}, {Delabrouille}, {Delouis}, {Di
  Valentino}, {Diego}, {Dor{\'e}}, {Douspis}, {Ducout}, {Dupac}, {Dusini},
  {Efstathiou}, {Elsner}, {En{\ss}lin}, {Eriksen}, {Fantaye}, {Farhang},
  {Fergusson}, {Fernandez-Cobos}, {Finelli}, {Forastieri}, {Frailis},
  {Franceschi}, {Frolov}, {Galeotta}, {Galli}, {Ganga}, {G{\'e}nova-Santos},
  {Gerbino}, {Ghosh}, {Gonz{\'a}lez-Nuevo}, {G{\'o}rski}, {Gratton},
  {Gruppuso}, {Gudmundsson}, {Hamann}, {Handley}, {Herranz}, {Hivon}, {Huang},
  {Jaffe}, {Jones}, {Karakci}, {Keih{\"a}nen}, {Keskitalo}, {Kiiveri}, {Kim},
  {Kisner}, {Knox}, {Krachmalnicoff}, {Kunz}, {Kurki-Suonio}, {Lagache},
  {Lamarre}, {Lasenby}, {Lattanzi}, {Lawrence}, {Le Jeune}, {Lemos},
  {Lesgourgues}, {Levrier}, {Lewis}, {Liguori}, {Lilje}, {Lilley}, {Lindholm},
  {L{\'o}pez-Caniego}, {Lubin}, {Ma}, {Mac{\'\i}as-P{\'e}rez}, {Maggio},
  {Maino}, {Mandolesi}, {Mangilli}, {Marcos-Caballero}, {Maris}, {Martin},
  {Martinelli}, {Mart{\'\i}nez- Gonz{\'a}lez}, {Matarrese}, {Mauri}, {McEwen},
  {Meinhold}, {Melchiorri}, {Mennella}, {Migliaccio}, {Millea}, {Mitra},
  {Miville-Desch{\^e}nes}, {Molinari}, {Montier}, {Morgante}, {Moss}, {Natoli},
  {N{\o}rgaard-Nielsen}, {Pagano}, {Paoletti}, {Partridge}, {Patanchon},
  {Peiris}, {Perrotta}, {Pettorino}, {Piacentini}, {Polastri}, {Polenta},
  {Puget}, {Rachen}, {Reinecke}, {Remazeilles}, {Renzi}, {Rocha}, {Rosset},
  {Roudier}, {Rubi{\~n}o-Mart{\'\i}n}, {Ruiz-Granados}, {Salvati}, {Sandri},
  {Savelainen}, {Scott}, {Shellard}, {Sirignano}, {Sirri}, {Spencer},
  {Sunyaev}, {Suur-Uski}, {Tauber}, {Tavagnacco}, {Tenti}, {Toffolatti},
  {Tomasi}, {Trombetti}, {Valenziano}, {Valiviita}, {Van Tent}, {Vibert},
  {Vielva}, {Villa}, {Vittorio}, {Wandelt}, {Wehus}, {White}, {White},
  {Zacchei}, \& {Zonca}}]{Aghanim2018_Planck_VI_Cosmological_parameters}
{Planck Collaboration}, {Aghanim}, N., {Akrami}, Y., {et~al.} 2018, arXiv
  e-prints, arXiv:1807.06209

\bibitem[{{Planck Collaboration} {et~al.}(2016{\natexlab{b}}){Planck
  Collaboration}, {Aghanim}, {Ashdown}, {Aumont}, {Baccigalupi}, {Ballardini},
  {Band ay}, {Barreiro}, {Bartolo}, {Basak}, {Benabed}, {Bernard},
  {Bersanelli}, {Bielewicz}, {Bonavera}, {Bond}, {Borrill}, {Bouchet},
  {Boulanger}, {Burigana}, {Calabrese}, {Cardoso}, {Carron}, {Chiang},
  {Colombo}, {Comis}, {Couchot}, {Coulais}, {Crill}, {Curto}, {Cuttaia}, {de
  Bernardis}, {de Zotti}, {Delabrouille}, {Di Valentino}, {Dickinson}, {Diego},
  {Dor{\'e}}, {Douspis}, {Ducout}, {Dupac}, {Dusini}, {Elsner}, {En{\ss}lin},
  {Eriksen}, {Falgarone}, {Fantaye}, {Finelli}, {Forastieri}, {Frailis},
  {Fraisse}, {Franceschi}, {Frolov}, {Galeotta}, {Galli}, {Ganga},
  {G{\'e}nova-Santos}, {Gerbino}, {Ghosh}, {Giraud-H{\'e}raud},
  {Gonz{\'a}lez-Nuevo}, {G{\'o}rski}, {Gruppuso}, {Gudmundsson}, {Hansen},
  {Helou}, {Henrot-Versill{\'e}}, {Herranz}, {Hivon}, {Huang}, {Jaffe},
  {Jones}, {Keih{\"a}nen}, {Keskitalo}, {Kiiveri}, {Kisner}, {Krachmalnicoff},
  {Kunz}, {Kurki-Suonio}, {Lamarre}, {Langer}, {Lasenby}, {Lattanzi},
  {Lawrence}, {Le Jeune}, {Levrier}, {Lilje}, {Lilley}, {Lindholm},
  {L{\'o}pez-Caniego}, {Ma}, {Mac{\'\i}as-P{\'e}rez}, {Maggio}, {Maino}, {Mand
  olesi}, {Mangilli}, {Maris}, {Martin}, {Mart{\'\i}nez-Gonz{\'a}lez},
  {Matarrese}, {Mauri}, {McEwen}, {Melchiorri}, {Mennella}, {Migliaccio},
  {Miville-Desch{\^e}nes}, {Molinari}, {Moneti}, {Montier}, {Morgante}, {Moss},
  {Natoli}, {Oxborrow}, {Pagano}, {Paoletti}, {Patanchon}, {Perdereau},
  {Perotto}, {Pettorino}, {Piacentini}, {Plaszczynski}, {Polastri}, {Polenta},
  {Puget}, {Rachen}, {Racine}, {Reinecke}, {Remazeilles}, {Renzi}, {Rocha},
  {Rosset}, {Rossetti}, {Roudier}, {Rubi{\~n}o-Mart{\'\i}n}, {Ruiz-Granados},
  {Salvati}, {Sandri}, {Savelainen}, {Scott}, {Sirignano}, {Sirri}, {Soler},
  {Spencer}, {Suur-Uski}, {Tauber}, {Tavagnacco}, {Tenti}, {Toffolatti},
  {Tomasi}, {Tristram}, {Trombetti}, {Valiviita}, {Van Tent}, {Vielva},
  {Villa}, {Vittorio}, {Wandelt}, {Wehus}, {Zacchei}, \&
  {Zonca}}]{Aghanim2016_Planck_XLVIII_Dust_emission}
{Planck Collaboration}, {Aghanim}, N., {Ashdown}, M., {et~al.}
  2016{\natexlab{b}}, \aap, 596, A109

\bibitem[{{Pozzetti} {et~al.}(2010){Pozzetti}, {Bolzonella}, {Zucca},
  {Zamorani}, {Lilly}, {Renzini}, {Moresco}, {Mignoli}, {Cassata}, {Tasca},
  {Lamareille}, {Maier}, {Meneux}, {Halliday}, {Oesch}, {Vergani}, {Caputi},
  {Kova{\v c}}, {Cimatti}, {Cucciati}, {Iovino}, {Peng}, {Carollo}, {Contini},
  {Kneib}, {Le F{\'e}vre}, {Mainieri}, {Scodeggio}, {Bardelli}, {Bongiorno},
  {Coppa}, {de la Torre}, {de Ravel}, {Franzetti}, {Garilli}, {Kampczyk},
  {Knobel}, {Le Borgne}, {Le Brun}, {Pell{\`o}}, {Perez Montero},
  {Ricciardelli}, {Silverman}, {Tanaka}, {Tresse}, {Abbas}, {Bottini}, {Cappi},
  {Guzzo}, {Koekemoer}, {Leauthaud}, {Maccagni}, {Marinoni}, {McCracken},
  {Memeo}, {Porciani}, {Scaramella}, {Scarlata}, \& {Scoville}}]{Pozzetti2010}
{Pozzetti}, L., {Bolzonella}, M., {Zucca}, E., {et~al.} 2010, \aap, 523, A13

\bibitem[{{Primack} {et~al.}(2011){Primack}, {Dom{\'{\i}}nguez}, {Gilmore}, \&
  {Somerville}}]{Primack2011}
{Primack}, J.~R., {Dom{\'{\i}}nguez}, A., {Gilmore}, R.~C., \& {Somerville},
  R.~S. 2011, in American Institute of Physics Conference Series, Vol. 1381,
  American Institute of Physics Conference Series, ed. F.~A. {Aharonian},
  W.~{Hofmann}, \& F.~M. {Rieger}, 72--83

\bibitem[{{Prochaska} \& {Herbert-Fort}(2004)}]{Prochaska_Herbert-Fort2004}
{Prochaska}, J.~X. \& {Herbert-Fort}, S. 2004, \pasp, 116, 622

\bibitem[{{Qi} {et~al.}(2019){Qi}, {Cao}, {Pan}, \& {Li}}]{Qi2019}
{Qi}, J.-Z., {Cao}, S., {Pan}, Y., \& {Li}, J. 2019, Physics of the Dark
  Universe, 26, 100338

\bibitem[{{Rachford} {et~al.}(2002){Rachford}, {Snow}, {Tumlinson}, {Shull},
  {Blair}, {Ferlet}, {Friedman}, {Gry}, {Jenkins}, {Morton}, {Savage},
  {Sonnentrucker}, {Vidal-Madjar}, {Welty}, \& {York}}]{Rachford2002}
{Rachford}, B.~L., {Snow}, T.~P., {Tumlinson}, J., {et~al.} 2002, \apj, 577,
  221

\bibitem[{{Rao} {et~al.}(2006){Rao}, {Turnshek}, \& {Nestor}}]{Rao2006}
{Rao}, S.~M., {Turnshek}, D.~A., \& {Nestor}, D.~B. 2006, \apj, 636, 610

\bibitem[{{Rauch}(1998)}]{Rauch1998}
{Rauch}, M. 1998, \araa, 36, 267

\bibitem[{{Reddy} {et~al.}(2012){Reddy}, {Dickinson}, {Elbaz}, {Morrison},
  {Giavalisco}, {Ivison}, {Papovich}, {Scott}, {Buat}, {Burgarella},
  {Charmandaris}, {Daddi}, {Magdis}, {Murphy}, {Altieri}, {Aussel},
  {Dannerbauer}, {Dasyra}, {Hwang}, {Kartaltepe}, {Leiton}, {Magnelli}, \&
  {Popesso}}]{Reddy2012}
{Reddy}, N., {Dickinson}, M., {Elbaz}, D., {et~al.} 2012, \apj, 744, 154

\bibitem[{{Reddy} \& {Steidel}(2009)}]{Reddy2009}
{Reddy}, N.~A. \& {Steidel}, C.~C. 2009, \apj, 692, 778

\bibitem[{{Riess} {et~al.}(2018){Riess}, {Casertano}, {Yuan}, {Macri},
  {Bucciarelli}, {Lattanzi}, {MacKenty}, {Bowers}, {Zheng}, {Filippenko},
  {Huang}, \& {Anderson}}]{Riess2018}
{Riess}, A.~G., {Casertano}, S., {Yuan}, W., {et~al.} 2018, \apj, 861, 126

\bibitem[{{Riess} {et~al.}(1998){Riess}, {Filippenko}, {Challis},
  {Clocchiatti}, {Diercks}, {Garnavich}, {Gilliland}, {Hogan}, {Jha},
  {Kirshner}, {Leibundgut}, {Phillips}, {Reiss}, {Schmidt}, {Schommer},
  {Smith}, {Spyromilio}, {Stubbs}, {Suntzeff}, \& {Tonry}}]{Riess1998}
{Riess}, A.~G., {Filippenko}, A.~V., {Challis}, P., {et~al.} 1998, \aj, 116,
  1009

\bibitem[{{Riess} {et~al.}(2011){Riess}, {Macri}, {Casertano}, {Lampeitl},
  {Ferguson}, {Filippenko}, {Jha}, {Li}, \& {Chornock}}]{Riess2011}
{Riess}, A.~G., {Macri}, L., {Casertano}, S., {et~al.} 2011, \apj, 730, 119

\bibitem[{{Riess} {et~al.}(2016){Riess}, {Macri}, {Hoffmann}, {Scolnic},
  {Casertano}, {Filippenko}, {Tucker}, {Reid}, {Jones}, {Silverman},
  {Chornock}, {Challis}, {Yuan}, {Brown}, \& {Foley}}]{Riess2016}
{Riess}, A.~G., {Macri}, L.~M., {Hoffmann}, S.~L., {et~al.} 2016, \apj, 826, 56

\bibitem[{{Ryan-Weber} {et~al.}(2006){Ryan-Weber}, {Pettini}, \&
  {Madau}}]{Ryan-Weber2006}
{Ryan-Weber}, E.~V., {Pettini}, M., \& {Madau}, P. 2006, \mnras, 371, L78

\bibitem[{{Ryden}(2016)}]{Ryden2016}
{Ryden}, B. 2016, {Introduction to Cosmology}

\bibitem[{{Sakstein} \& {Jain}(2017)}]{Sakstein_Jain2017}
{Sakstein}, J. \& {Jain}, B. 2017, \prl, 119, 251303

\bibitem[{{Salmon} {et~al.}(2018){Salmon}, {Coe}, {Bradley}, {Brada{\v{c}}},
  {Strait}, {Paterno-Mahler}, {Huang}, {Oesch}, {Zitrin}, {Acebron}, {Cibirka},
  {Kikuchihara}, {Oguri}, {Brammer}, {Sharon}, {Trenti}, {Avila}, {Ogaz},
  {Andrade-Santos}, {Carrasco}, {Cerny}, {Dawson}, {Frye}, {Hoag}, {Jones},
  {Mainali}, {Ouchi}, {Rodney}, {Stark}, \& {Umetsu}}]{Salmon2018}
{Salmon}, B., {Coe}, D., {Bradley}, L., {et~al.} 2018, \apj, 864, L22

\bibitem[{{Schechter}(1976)}]{Schechter1976}
{Schechter}, P. 1976, \apj, 203, 297

\bibitem[{{Schiminovich} {et~al.}(2005){Schiminovich}, {Ilbert}, {Arnouts},
  {Milliard}, {Tresse}, {Le F{\`e}vre}, {Treyer}, {Wyder}, {Budav{\'a}ri},
  {Zucca}, {Zamorani}, {Martin}, {Adami}, {Arnaboldi}, {Bardelli}, {Barlow},
  {Bianchi}, {Bolzonella}, {Bottini}, {Byun}, {Cappi}, {Contini}, {Charlot},
  {Donas}, {Forster}, {Foucaud}, {Franzetti}, {Friedman}, {Garilli},
  {Gavignaud}, {Guzzo}, {Heckman}, {Hoopes}, {Iovino}, {Jelinsky}, {Le Brun},
  {Lee}, {Maccagni}, {Madore}, {Malina}, {Marano}, {Marinoni}, {McCracken},
  {Mazure}, {Meneux}, {Morrissey}, {Neff}, {Paltani}, {Pell{\`o}}, {Picat},
  {Pollo}, {Pozzetti}, {Radovich}, {Rich}, {Scaramella}, {Scodeggio},
  {Seibert}, {Siegmund}, {Small}, {Szalay}, {Vettolani}, {Welsh}, {Xu}, \&
  {Zanichelli}}]{Schiminovich2005}
{Schiminovich}, D., {Ilbert}, O., {Arnouts}, S., {et~al.} 2005, \apjl, 619, L47

\bibitem[{{Schlegel} {et~al.}(1998){Schlegel}, {Finkbeiner}, \&
  {Davis}}]{Schlegel1998}
{Schlegel}, D.~J., {Finkbeiner}, D.~P., \& {Davis}, M. 1998, \apj, 500, 525

\bibitem[{{Segers} {et~al.}(2016){Segers}, {Crain}, {Schaye}, {Bower},
  {Furlong}, {Schaller}, \& {Theuns}}]{Segers2016}
{Segers}, M.~C., {Crain}, R.~A., {Schaye}, J., {et~al.} 2016, \mnras, 456, 1235

\bibitem[{{Songaila}(2001)}]{Songaila2001}
{Songaila}, A. 2001, \apjl, 561, L153

\bibitem[{{Songaila} \& {Cowie}(2010)}]{Songaila_Cowie2010}
{Songaila}, A. \& {Cowie}, L.~L. 2010, \apj, 721, 1448

\bibitem[{{Suyu} {et~al.}(2013){Suyu}, {Auger}, {Hilbert}, {Marshall}, {Tewes},
  {Treu}, {Fassnacht}, {Koopmans}, {Sluse}, {Blandford}, {Courbin}, \&
  {Meylan}}]{Suyu2013}
{Suyu}, S.~H., {Auger}, M.~W., {Hilbert}, S., {et~al.} 2013, \apj, 766, 70

\bibitem[{{Tejos} {et~al.}(2012){Tejos}, {Morris}, {Crighton}, {Theuns},
  {Altay}, \& {Finn}}]{Tejos2012}
{Tejos}, N., {Morris}, S.~L., {Crighton}, N. H.~M., {et~al.} 2012, \mnras, 425,
  245

\bibitem[{{Tejos} {et~al.}(2014){Tejos}, {Morris}, {Finn}, {Crighton},
  {Bechtold}, {Jannuzi}, {Schaye}, {Theuns}, {Altay}, {Le F{\`e}vre},
  {Ryan-Weber}, \& {Dav{\'e}}}]{Tejos2014}
{Tejos}, N., {Morris}, S.~L., {Finn}, C.~W., {et~al.} 2014, \mnras, 437, 2017

\bibitem[{{Tumlinson} {et~al.}(2017){Tumlinson}, {Peeples}, \&
  {Werk}}]{Tumlinson2017}
{Tumlinson}, J., {Peeples}, M.~S., \& {Werk}, J.~K. 2017, \araa, 55, 389

\bibitem[{{Vavry{\v c}uk}(2017{\natexlab{a}})}]{Vavrycuk2017b}
{Vavry{\v c}uk}, V. 2017{\natexlab{a}}, \mnras, 470, L44

\bibitem[{{Vavry{\v c}uk}(2017{\natexlab{b}})}]{Vavrycuk2017a}
{Vavry{\v c}uk}, V. 2017{\natexlab{b}}, \mnras, 465, 1532

\bibitem[{{Vavry{\v{c}}uk}(2018)}]{Vavrycuk2018}
{Vavry{\v{c}}uk}, V. 2018, \mnras, 478, 283

\bibitem[{{Vavry{\v{c}}uk}(2019)}]{Vavrycuk2019}
{Vavry{\v{c}}uk}, V. 2019, \mnras, 489, L63

\bibitem[{{Vavry{\v{c}}uk} \& {Kroupa}(2020)}]{Vavrycuk_Kroupa2020}
{Vavry{\v{c}}uk}, V. \& {Kroupa}, P. 2020, \mnras, 497, 378

\bibitem[{{Venemans} {et~al.}(2017){Venemans}, {Walter}, {Decarli},
  {Ba{\~n}ados}, {Carilli}, {Winters}, {Schuster}, {da Cunha}, {Fan}, {Farina},
  {Mazzucchelli}, {Rix}, \& {Weiss}}]{Venemans2017}
{Venemans}, B.~P., {Walter}, F., {Decarli}, R., {et~al.} 2017, \apj, 851, L8

\bibitem[{{Vielva} {et~al.}(2004){Vielva}, {Mart{\'{\i}}nez-Gonz{\'a}lez},
  {Barreiro}, {Sanz}, \& {Cay{\'o}n}}]{Vielva2004}
{Vielva}, P., {Mart{\'{\i}}nez-Gonz{\'a}lez}, E., {Barreiro}, R.~B., {Sanz},
  J.~L., \& {Cay{\'o}n}, L. 2004, \apj, 609, 22

\bibitem[{{Vitale} \& {Chen}(2018)}]{Vitale_Chen2018}
{Vitale}, S. \& {Chen}, H.-Y. 2018, \prl, 121, 021303

\bibitem[{{Wakker} {et~al.}(2015){Wakker}, {Hernandez}, {French}, {Kim},
  {Oppenheimer}, \& {Savage}}]{Wakker2015}
{Wakker}, B.~P., {Hernandez}, A.~K., {French}, D.~M., {et~al.} 2015, \apj, 814,
  40

\bibitem[{{Watson} {et~al.}(2015){Watson}, {Christensen}, {Knudsen}, {Richard},
  {Gallazzi}, \& {Micha{\l}owski}}]{Watson2015}
{Watson}, D., {Christensen}, L., {Knudsen}, K.~K., {et~al.} 2015, \nat, 519,
  327

\bibitem[{{Wei} \& {Wu}(2017)}]{Wei_Wu2017}
{Wei}, J.-J. \& {Wu}, X.-F. 2017, \apj, 838, 160

\bibitem[{{Weinberg} {et~al.}(2013){Weinberg}, {Mortonson}, {Eisenstein},
  {Hirata}, {Riess}, \& {Rozo}}]{Weinberg2013}
{Weinberg}, D.~H., {Mortonson}, M.~J., {Eisenstein}, D.~J., {et~al.} 2013,
  \physrep, 530, 87

\bibitem[{{Weingartner} \& {Draine}(2001)}]{Weingartner_Draine2001}
{Weingartner}, J.~C. \& {Draine}, B.~T. 2001, \apj, 548, 296

\bibitem[{{Wolfe} {et~al.}(2005){Wolfe}, {Gawiser}, \& {Prochaska}}]{Wolfe2005}
{Wolfe}, A.~M., {Gawiser}, E., \& {Prochaska}, J.~X. 2005, \araa, 43, 861

\bibitem[{{Wright}(1982)}]{Wright1982}
{Wright}, E.~L. 1982, \apj, 255, 401

\bibitem[{{Wright}(1987)}]{Wright1987}
{Wright}, E.~L. 1987, \apj, 320, 818

\bibitem[{{Xie} {et~al.}(2016){Xie}, {Shao}, {Shen}, {Liu}, \& {Li}}]{Xie2016}
{Xie}, X., {Shao}, Z., {Shen}, S., {Liu}, H., \& {Li}, L. 2016, \apj, 824, 38

\bibitem[{{Xie} {et~al.}(2015){Xie}, {Shen}, {Shao}, \& {Yin}}]{Xie2015}
{Xie}, X., {Shen}, S., {Shao}, Z., \& {Yin}, J. 2015, \apjl, 802, L16

\bibitem[{{Yabe} {et~al.}(2009){Yabe}, {Ohta}, {Iwata}, {Sawicki}, {Tamura},
  {Akiyama}, \& {Aoki}}]{Yabe2009}
{Yabe}, K., {Ohta}, K., {Iwata}, I., {et~al.} 2009, \apj, 693, 507

\bibitem[{{Yu} \& {Wang}(2016)}]{Yu_Wang2016}
{Yu}, H. \& {Wang}, F.~Y. 2016, \apj, 828, 85

\bibitem[{{Zavala} {et~al.}(2015){Zavala}, {Micha{\l}owski}, {Aretxaga},
  {Wilson}, {Hughes}, {Monta{\~n}a}, {Dunlop}, {Pope},
  {S{\'a}nchez-Arg{\"u}elles}, {Yun}, \& {Zeballos}}]{Zavala2015}
{Zavala}, J.~A., {Micha{\l}owski}, M.~J., {Aretxaga}, I., {et~al.} 2015,
  \mnras, 453, L88

\bibitem[{{Zinnecker} \& {Yorke}(2007)}]{Zinnecker_Yorke2007}
{Zinnecker}, H. \& {Yorke}, H.~W. 2007, Annual Review of Astronomy and
  Astrophysics, 45, 481

\bibitem[{{Zwaan} {et~al.}(2005){Zwaan}, {van der Hulst}, {Briggs},
  {Verheijen}, \& {Ryan-Weber}}]{Zwaan2005}
{Zwaan}, M.~A., {van der Hulst}, J.~M., {Briggs}, F.~H., {Verheijen}, M.~A.~W.,
  \& {Ryan-Weber}, E.~V. 2005, \mnras, 364, 1467

\end{thebibliography}

\end{document}